\documentclass[useAMS,usenatbib]{mn2e}
\usepackage{color}
\usepackage{array} 
\usepackage{amsmath}
\usepackage{graphicx}
\usepackage{epstopdf}
\usepackage[labelfont=bf,labelsep=period,justification=raggedright,singlelinecheck=false,font=small]{caption}
\usepackage{wrapfig}
\usepackage{float}
\usepackage{amssymb}
\usepackage[normalem]{ulem}
\usepackage{hyperref}
\usepackage{undertilde}
\usepackage[T1]{fontenc}
\usepackage[utf8]{inputenc}
\usepackage{comment}
\usepackage[export]{adjustbox}
\usepackage[colorinlistoftodos]{todonotes}
\usepackage{scalerel}
\usepackage{tikz}
\usetikzlibrary{svg.path}

\definecolor{orcidlogocol}{HTML}{A6CE39}
\tikzset{
  orcidlogo/.pic={
    \fill[orcidlogocol] svg{M256,128c0,70.7-57.3,128-128,128C57.3,256,0,198.7,0,128C0,57.3,57.3,0,128,0C198.7,0,256,57.3,256,128z};
    \fill[white] svg{M86.3,186.2H70.9V79.1h15.4v48.4V186.2z}
                 svg{M108.9,79.1h41.6c39.6,0,57,28.3,57,53.6c0,27.5-21.5,53.6-56.8,53.6h-41.8V79.1z M124.3,172.4h24.5c34.9,0,42.9-26.5,42.9-39.7c0-21.5-13.7-39.7-43.7-39.7h-23.7V172.4z}
                 svg{M88.7,56.8c0,5.5-4.5,10.1-10.1,10.1c-5.6,0-10.1-4.6-10.1-10.1c0-5.6,4.5-10.1,10.1-10.1C84.2,46.7,88.7,51.3,88.7,56.8z};
  }
}

\newcommand\orcidicon[1]{\href{https://orcid.org/#1}{\mbox{\scalerel*{
\begin{tikzpicture}[yscale=-1,transform shape]
\pic{orcidlogo};
\end{tikzpicture}
}{|}}}}

\def\apj{ApJ}
\def\aap{A\& A}
\def\apjl{ApJL}
\def\apjs{ApJS}
\def\mnras{MNRAS}
\def\jcap{JCAP}

\def\aj{AJ}
\def\araa{ARA\& A}
\def\nat{Nature}

\newcommand{\X}{\hbox{$X_{\rm eff}$}}
\newcommand{\Xb}{\hbox{$X_{\rm b, eff}$}}
\renewcommand{\L}{\hbox{$\lambda_{\rm eff}$}}
\newcommand{\Lb}{\hbox{$\lambda_{\rm b, eff}$}}
\newcommand\altaffilmark[1]{$^{#1}$}
\newcommand\altaffiltext[1]{$^{#1}$}

\newcommand{\cddf}{{\sc cddf}}
\newcommand{\igm}{{\sc igm}}

\newcommand{\lls}{{\sc lls}}
\newcommand{\dla}{{\sc dla}}
\newcommand{\qso}{{\sc qso}}

\newcommand{\rt}{{\sc rt}}
\newcommand{\col}{{\sc colossus}}
\newcommand{\pdf}{{\sc pdf}}

\newcommand{\appropto}{\mathrel{\vcenter{
  \offinterlineskip\halign{\hfil$##$\cr\propto\cr\noalign{\kern2pt}\sim\cr\noalign{\kern-2pt}}}}}
    
\restylefloat{table}
\begin{document}
\defcitealias{Theuns21}{TT21}
\title[Lyman-limit systems]{A halo model for cosmological Lyman-limit systems}

\author[Theuns \& Chan]
  {Tom Theuns\altaffilmark{1}$^{\orcidicon{0000-0002-3790-9520}}$
  and T. K. ~Chan\altaffilmark{1,2}\thanks{Email: (TKC)tsang.k.chan@durham.ac.uk}$^{\orcidicon{0000-0003-2544-054X}}$ \\
  \altaffiltext{1}{Institute for Computational Cosmology, Department of Physics, Durham University, South Road, Durham DH1 3LE, UK}\\ 
  \altaffiltext{2}{Department of Astronomy and Astrophysics, the University of Chicago, Chicago, IL60637, USA}
}

\maketitle

\begin{abstract}
We present an analytical model for cosmological Lyman-limit systems (\lls) that successfully reproduces the observed evolution of the mean free path (\L) of ionizing photons. The evolution 
of the co-moving mean free path is predominantly a consequence of the changing meta galactic photo-ionization rate and the increase with cosmic time of the minimum mass below which halos lose their gas due to photo-heating. In the model, Lyman-limit absorption is caused by highly ionized gas in the outskirt of dark matter halos. We exploit the association with halos to compute statistical properties of \L\ and of the bias, $b$, of \lls. The latter increases from $b\sim 1.5\to 2.6$ from redshifts $z=2\to 6$. Combined with the rapid increase with redshift of the bias of the halos that host a quasar, the model predicts a rapid drop in the value of \L\ when {\em measured} in quasar spectra from $z=5\to 6$, whereas the {\em actual} value of \L\ falls more smoothly. We derive an expression for the effective optical depth due to Lyman limit absorption as a function of wavelength and show that it depends sensitively on the poorly constrained number density of \lls\ as a function of column density. The optical depth drops below unity for all wavelengths below a redshift of $\sim 2.5$ which is therefore the epoch when the Universe first became transparent to ionizing photons.
\end{abstract}


\label{firstpage}

\section{Introduction}
\label{sec:introduction}
Hydrogen in the intergalactic medium (hereafter \igm) is so highly ionized 
that it does not produce a significant Gunn-Peterson trough \citep{Gunn65} in quasar spectra below a redshift of $z\sim 6$ \citep{Fan06}. At higher redshifts, several independent observations suggest that the \igm\ may be significantly neutral. These include a detection of a damping wing in the spectra of $z>7$ quasars \citep{Mortlock11, Davies18} as well as other observations \cite[e.g.][]{Mason18}. The measurement of the Thompson-optical depth to the cosmic microwave background from free electrons also suggest that the Universe transitioned from mostly neutral to mostly ionized around $z\sim 7.5$ \citep{Planck20}. For reviews on the physics of the \igm\ and its connection to reionization, see e.g. \citet{Meiksin09} or \citet{McQuinn16},
and for a more observational perspective, see e.g. \citet{Rauch98}.

Even when the Universe is highly ionized on average, the remaining neutral hydrogen is sufficiently abundant to limit the distance that a typical ionizing photon can travel from its source before being absorbed.  This distance can be quantified either by the attenuation length $\lambda_{\rm eff}$, defined below in Eq.~(\ref{eq:lambda}), or the \lq mean free path\rq. The relation between these quantities is examined in more detail in Appendix \ref{sect:appendixB}. The attenuation length and the emissivity of ionizing sources together determine the amplitude of the ionizing background \citep[e.g.][]{Haardt96, Miralda03,McQuinn11,Faucher09,Haar12UVbackground}. 

The absorbers of ionizing photons are usually characterised in terms of their neutral hydrogen column density, $N_{\rm HI}$, and are labeled as \lq Lyman-$\alpha$ forest\rq\
($N_{\rm HI}< 10^{17.2}{\rm cm}^{-2}$), Lyman-limit systems (\lls's, $10^{17.2}{\rm cm}^{-2}\leq N_{\rm HI}< 10^{20.3}{\rm cm}^{-2}$), and damped Lyman-$\alpha$ absorbers
(\dla's, $N_{\rm HI}\geq 10^{20.3}{\rm cm}^{-2}$, see e.g. \citealt{Rauch98}). The optical depth of an ionizing photon with energy of 1 Rydberg is unity at $N_{\rm HI}=10^{17.2}{\rm cm}^{-2}$, whereas the Lyman-$\alpha$ line shows an obvious damping wing above a column density of $N_{\rm HI}=10^{20.3}{\rm cm}^{-2}$ - hence the labels. It is also common parlance to refer to absorbers with column density just below $10^{17.2}{\rm cm}^{-2}$ as sub-\lls's, and those close to but below the \dla\ threshold as super-\lls's or sub-\dla's.

The column-density distribution function (hereafter \cddf), is the number density of absorbers with a given value of $N_{\rm HI}$ (per unit co-moving path length, to be defined below), and the normalization and shape of this function sets \L. Sub-\lls's and super-\lls's together mostly determine the value of \L, because the numerous Lyman-$\alpha$ forest absorbers just have too low a column density to contribute significantly to \L, whereas the strongly absorbing \dla's are simply too rare. Unfortunately, it is difficult to measure accurately the column density of lines in the important range of $10^{16}-10^{20}{\rm cm}^{-2}$ because the whole Lyman-series of absorption lines associated with the absorber is partially or completely saturated. Estimates of \L\ then require extrapolating the \cddf\ in the \lls\ range, {\em i.e.} just that range of the \cddf\ that is the most important for accurately determining \L\ \citep[e.g.][]{Faucher09, Haar12UVbackground}.

\cite{Proc09opacity} suggested an alternative method for measuring \L, namely stacking quasar transmission spectra in bins of emission redshift and measuring the decrease in transmission caused by the ionization edge of the hydrogen atom - {\em i.e.} the reduction in transmission\footnote{After correction for the reduction in transmission caused by the Lyman series of absorption lines, and accounting for other observational effects.} of photons with energies $h\nu\geq 13.6{\rm eV}$ \citep[see also][]{Fumagalli13, Omeara13, Worseck14, Becker21}. The measured value of \L\ decreases rapidly with increasing $z$, approximately $\propto (1+z)^{-\eta}$ with $\eta\approx 5.4$, over the redshift range $z=2.3\to 5.5$ \citep{Worseck14}. This is much faster than would be the case if the absorbers had constant co-moving density and a constant proper cross-section, which would yield $\L\propto (1+z)^{-3}$, demonstrating that the absorbers evolve. Given that the (co-moving) number density of absorbers presumably {\em increases}  with cosmic time as structure grows, and that the intrinsic sizes of the absorbers presumably also grow with time, with both effects tending to {\em reduce} \L, one might naively expect that \L\ evolves {\em slower} than $(1+z)^{-3}$ - which is exactly opposite from what is observed.  
\cite{Prochaska10} discusses several possible reasons for this unexpected evolution, settling on the suggestion that it must be that absorbers become more highly ionized with decreasing $z$.

\cite{Becker21} uses the method of \cite{Proc09opacity} to measure \L\ in a set of $z\sim 6$ quasars. Accounting for the radiation of the quasar itself -- the proximity effect -- they infer a sharp drop in \L\ from $z=5\to 6$, much faster than an extrapolation of the $\propto (1+z)^{-\eta}$ would predict. They claim that this rapid change signals the transition from an ionized to a mostly neutral \igm, and hence claim that their measurements are probing the tail-end of the epoch of reionization 
\citep[see also][]{Gaikwad23}.

Numerical models to predict the evolution of \L\ are challenging, requiring radiative transfer (hereafter \rt) at high resolution to capture the transition from ionized to neutral gas with increasing density in a computational volume that is large enough to sample the relatively rare strong absorbers that set \L. \cite{Altay11} and \cite{McQuinn11} both post-processed simulations with \rt, showing that they can reproduce the observed \cddf, including the transition from Lyman-$\alpha$ forest to \dla's. These papers show that the \cddf\ evolves relatively slowly, in agreement with observations (see also \citealt{Rahmati13}). \cite{Altay13} further show that these predictions are insensitive to the uncertainties in the modelling  caused by {\em galaxy} formation ({\em i.e.} the implementation of feedback from massive stars and quasars), which only affects the \cddf\ at high column-densities, $N_{\rm HI}\ge 10^{21}{\rm cm}^{-2}$, that have little effect on \L. All modellers agree that cold gas, accreting onto halos, is the dominant contributor to \lls's \cite[e.g.][]{Altay11, Faucher11, Fumagalli11, Yajima12, vandevoort12, Rahmati13}. Analytic models
for the evolution of absorbers,  sometimes augmented with observational constraints or numerical models, are also discussed by \cite{Erkal15} and \cite{Muno16flatgamma}, and we will contrast our approach and results with theirs below.

A flurry of recent papers used simulations that include radiative transfer performed either on the fly or in post-processing to investigate the claim by \cite{Becker21} that \L\ drops sharply from $z=5\to 6$ \citep[e.g.][]{Keating20, Daloisio20, Cain21,  Garaldi22, Gaikwad23}. The authors attribute the drop to this redshift range probing the tail-end of reionization.
We will return to this issue in \S \ref{sect:b_L}.

In this paper we present an analytical model for the \cddf\ in the \lls\ and \dla\ range, making the assumption that (strong) absorption lines are caused by gas in halos. The model of absorbers and their connection to \L\ are presented in section~\ref{sect:halomodel}.
Section~\ref{sect:bias} discusses clustering of absorbers and the impact of bias on \L. We also show how the statistical properties of the attenuation relate to clustering of halos. Section~\ref{sect:measuring} exploits the model to compute the wavelength dependence of the optical depth, resulting in a new model for the combined effect of many \lls\ on the mean transmission. Section~5 summarizes our results.
We use the \cite{Planck15} values of cosmological parameters (final column of their table~4), Hubble parameter $h=0.673$, baryon and matter density in units of the critical density of $\Omega_b=0.02230/h^2$ and $\Omega_m=(0.1188 + 0.02230 )/h^2$, a Helium abundance by mass of $Y=0.24531$, and when applicable apply the high-z approximation for the Hubble constant at redshift $z$, $H(z)=H_0\,\Omega_m^{1/2}\,(1+z)^{3/2}$, with $H_0$ the Hubble constant at $z=0$.
\section{The attenuation length in the halo model}
\label{sect:halomodel}
We begin this section by briefly reviewing the relation between the attenuation length, \L, and the column density distribution function ($f(N_{\rm H})$, hereafter \cddf). We then extend the model of \cite{Theuns21} of \dla's to 
the lower column density \lls\ (\S~\ref{sect:cddf}), and use the resulting \cddf\ to derive the evolution of \L\ which we compare to observations. We infer the main drivers of the evolution of \L\ by varying the parameters that determine the \cddf\  (\S~\ref{sect:L}). We finish this section by comparing to the observed evolution of \lls's
(\S~\ref{sect:lls}).

\subsection{Relating \L\ to the \cddf}
\label{sect:intro}
Absorption of ionizing photons in the clumpy Universe occurs predominantly in approximately discrete \lq absorbers\rq\ with a range of neutral hydrogen column densities, $N_{\rm HI}$. Provided that these absorbers are Poisson distributed along a sight line ({\em i.e.}, provided we neglect any spatial correlations of absorbers: we account for clustering later on), the effective optical depth, $\tau_{\rm eff}$, per unit proper sight line distance, $dl$, at the Lyman limit, is \cite[e.g.][]{Paresce80, Meiksin93}
\begin{equation}
    \frac{d\tau_{\rm eff}}{dl} = \int_0^\infty\,\frac{d^2 N}{d l\,d N_{\rm HI}}\,\left[1-\exp(-\tau)\right]\,dN_{\rm HI}\,.
    \label{eq:dtau}
\end{equation}
Here, $N$ is the number of absorbers with column density $N_{\rm HI}$ per unit proper distance $dl$ and $\tau=\sigma_{\rm th}N_{\rm HI}$ is the optical depth of an absorber; $\sigma_{\rm th}$ is the photo-ionization cross section at the Lyman limit. We will be more careful about the wavelength dependence of this relation later on.

The attenuation length is usually expressed as a proper distance. To expose better the underlying physics, it is useful to separate the contributions to the evolution of \L\ that result from the expansion of the Universe and those that result from changes in the intrinsic properties of the absorbers. To enable this, \cite{Bahcall69} defined the dimensionless co-moving path length, $dX$, as 
\begin{align}
    dX \equiv \frac{H_0\,(1+z)^2}{H(z)}\,dz\equiv \frac{H_0\,(1+z)^3}{c}\,dl\,.
\label{eq:dX}
\end{align}
We note that $dX$ is not simply the co-moving analogue of the proper path length  $dl$. Combining the above relations yields
\begin{align}
 \frac{d\tau_{\rm eff}}{dX}(z) &= \int_0^\infty\,f(N_{\rm HI};z)\,\left[1-\exp(-\tau)\right]\,dN_{\rm HI}\,,
    \label{eq:dtaudl} 
\end{align}
where
\begin{align}
   f(N_{\rm HI};z) \equiv  \frac{d^2 N}{d X\,d N_{\rm HI}}(z)\,,
\end{align}
is now the number $N$ of absorbers with a given column density per $dX$, {\em i.e.} the \cddf\ at redshift $z$. 

The intensity of a beam of photons with frequency\footnote{Where $h\nu_{\rm th}=1$~Ryd is the binding energy of HI. We discuss the frequency dependence in more detail in \S~4. (From the context it should be clear when $h$ is Planck's constant and when it refers to the Hubble parameter.)} $\nu_{\rm th}$ travelling a co-moving path length $dX$ will be attenuated by a factor $\exp(-\tau_{\rm eff})$ on average. Setting $\tau_{\rm eff}=1$ in Eq.~(\ref{eq:dtaudl}) defines the {\em attenuation length}, \X,
\begin{align}
    \X(z) &= \left\{\int_0^\infty\,f(N_{\rm HI};z)\,[1-\exp(-\tau)]\,{\rm d}N_{\rm HI}\right\}^{-1}\,.
    \label{eq:X}
\end{align}
Any evolution of \X\ is due to the evolution of the \cddf, {\em i.e.} due to the evolution of the absorbers, rather than simply due to the expansion of the Universe. 

Finally, the proper attenuation length, \L, is related to \X\ by
\begin{align}
    \L &= \frac{c}{H_0\,(1+z)^3}\,\X\,.
    \label{eq:atp}
\end{align}

The quantity \L\ is sometimes referred to as mean free path. However, \L\ and mean free path are different characterizations of absorption, and are generally not numerically equal, as we demonstrate in Appendix \ref{sect:appendixB}

We continue by briefly reviewing the model of  \citet[][]{Theuns21} for strong {\sc HI} absorbers, which we extend to lower values of the column density to compute the evolution of the \cddf\ and hence that of the attenuation length.

\subsection{A model for the \cddf\ and its evolution}
\label{sect:cddf}

The model for strong {\sc Hi} absorbers by \citet[][hereafter TT21, see also \citealt{Erkal15}]{Theuns21} is based on the following two main approximations:
\begin{itemize}
    \item[({\rm i})] Gas in halos is spherically symmetrically distributed around the halo's centre of mass with a power-law density profile, $\rho(R)\propto R^{-2}$.
    \item[({\rm ii})] This gas is in photo-ionization equilibrium with the ionising background, and the neutral fraction can be estimated using approximate radiative transfer of ionizing photons penetrating radially inwards.
\end{itemize}
Spherical symmetry is clearly an approximation and it would be interesting to
investigate how sensitive the model's predictions depend on this approximation.

In more detail, we assume the gas density profile to be\footnote{The model neglects the $\sim 10$~per cent effect of helium. It is straightforward to generate the model described here for a different exponent of the radial profile, which may be a better fit to profiles measured in simulations, in particular at higher $z\gtrapprox 6$.}
\begin{align}
    n_{\rm H}(R)=n_{\rm H,h}\times\left(\frac{R_h}{R}\right)^2\,,
    \label{eq:nH}
\end{align}
Here, $n_{\rm H}(R)$ is the hydrogen density by number at distance $R$ from the centre of the halo, $R_h$ is the virial radius of that halo, and $n_{\rm H,h}$ is the density at $R_h$ ($n_{\rm H,h}=200\,f_{\rm gas}\,\langle n_{\rm H}\rangle\,/3$, with $\langle n_{\rm H}\rangle$ the cosmic mean hydrogen density and $f_{\rm gas}$, which is of order unity,  the gas fraction at $R_h$ in units of the cosmic mean). 
All these variables are in proper units. Assuming further that this halo is illuminated by an ionizing background characterised by a photo-ionization rate $\Gamma_0$, \citetalias{Theuns21} performs simplified radiative transfer to compute the neutral fraction, $x\equiv n_{\rm HI}/n_{\rm H}$ as a function of radius, assuming the gas is isothermal at a temperature of $T=10^4$~K.  Calculating numerically the optical depth $\tau$ at radius $R$ due to neutral gas between $R$ and $R_h$ yields the factor $\exp(-\tau)$ by which the photo-ionization rate at $R$ is suppressed compared to its value at $R_h$. As $\tau$ increases, the neutral fraction rises rapidly once $\tau\ge 1$, and the gas transitions from highly ionized to mostly neutral. 

In this paper we extend \citetalias{Theuns21}'s model in two ways: 
\begin{itemize}
    \item[(1)]we extrapolate the profile of Eq.~(\ref{eq:nH}) to values $>R_h$,
    \item[(2)]we no longer assume that the gas is isothermal at a temperature of $T=10^4{\rm K}$. 
\end{itemize}
In practice, we extrapolate Eq.~(\ref{eq:atp}) out to $R=8\,R_h$. This extrapolation allows us to compute the number of absorbers at column densities far below that of \dla's, and we will show that the predicted number of such absorbers agrees fairly well with observations. The reason to make changes to the gas temperature as well is as follows. At lower densities where the gas is highly ionized, the gas temperature is closer to $T\sim 1.5\times 10^{4}{\rm K}$ \cite[e.g.][]{Schaye00} at the redshifts of interest ({\em i.e.} $z=2\to 6$), and hence we would like to use this more realistic value for $T$. Choosing this higher temperature changes the neutral fraction due to the $T$-dependence of the recombination rate, at higher density it further changes the neutral fraction due to collisional ionization. To avoid that our self-shielded gas is affected by collisional ionizations, we want to keep the temperature of this gas at $T=10^4{\rm K}$. We therefore interpolate $T$ from $1.5\times 10^{4}{\rm K}$ at $\tau<1$ to $T=10^4\,{\rm K}$ at $\tau\ge 1.5$. Given that these changes are relatively minor, we continue to refer to this improved model as \lq \citetalias{Theuns21}\rq.

We show below that the \lls's that set \L\ are mostly highly ionized, and so even neglecting any self-shielding has little impact on our results. Given this, we make an even more simplified model in this paper which assumes that
gas in \lls's is optically thin. The motivation for making this approximation is twofold: ({\em i}) it dramatically simplifies the equations, and ({\em ii}) the attenuation length is nearly identical to that of the more accurate model. The simpler analytical expressions greatly clarify the relation between the evolution of \X\ and that of halos. The reason for ({\em ii}) is that most of the absorption is due to \lls\ which occur in highly ionized gas that is well described by the approximate model. The approximation does {\em not} capture the transition from \lls's to \dla's. We will refer to the more accurate model as \lq TT21\rq\, and to the model that makes the optically thin approximation as \lq the optically thin\rq\ model.

The neutral fraction of the gas with the density profile of Eq.~(\ref{eq:nH}) can be computed analytically in the optically thin model.
This also allows us to obtain an analytical expression for the column density along a sight line at impact parameter $b$,
\begin{align}
    N_{\rm HI}(b) &= N_{\rm HI, h}\times\left(\frac{R_h}{b}\right)^3\nonumber\\
N_{\rm HI, h} &\equiv  \frac{2\alpha_B}{\Gamma_0}\,n^2_{\rm H,h}\,R_h\int_0^\infty\,\frac{dq}{(1+q^2)^2}\nonumber\\
&=10^{15.5}\,{\rm cm}^{-2}\left(\frac{1+z}{4}\right)^5\,
\left(\frac{M_h}{10^{10}M_\odot}\right)^{1/3}\nonumber\\
&\times 
\left(\frac{f_{\rm gas}}{0.6}\right)^2\,
\left(\frac{\alpha_B(T)}{\alpha_B(1.5\times 10^4{\rm K})}\right)\,\left(\frac{\Gamma_0}{10^{-12}{\rm s}^{-1}}\right)^{-1}\,.
\label{eq:NHh}    
\end{align}
Here, $\alpha_B(T)$ is the case-B recombination coefficient, $T$ is the temperature of the gas, and $M_h$ is the virial mass of the halo. The value of $N_{\rm HI, h}$ assumes that the $1/R^2$ profile of the halo extends to infinity, {\em i.e.} there is a (relatively small) contribution to $N_{\rm HI}$ from gas outside the halo.

The column-density-weighted neutral fraction of the gas along a sight line is
\begin{align}
    \langle x\rangle &= \frac{\int_0^\infty x\,n_{\rm HI}\,dl}{\int_0^\infty n_{\rm HI} dl}=\frac{3x_h}{4}\,\left(\frac{N_{\rm HI}}{N_{\rm H,h}}\right)^{2/3}\,,
\end{align}
where $x_h=\alpha_{\rm B}n_{\rm H,h}/\Gamma_0$ is the neutral fraction at $R_h$, and $l$ is the path length. For $z=3$ and $\Gamma_0=10^{-12}{\rm s}^{-1}$, $x_h\approx 10^{-4}$, which means that $\langle x\rangle<x_m$ provided that $N_{\rm HI}<N_{\rm H,h}\,(x_m/x_h)^{3/2}$ or
$N_{\rm HI}\lessapprox 500\,N_{\rm H,h}$ taking $x_m=10^{-2}$. Comparing to Eq.~(\ref{eq:NHh}) then shows that up to columns of order a few times $10^{18}{\rm cm}^{-2}$, the absorbing gas is indeed highly ionized, $x<10^{-2}$, hence making the optically thin approximation is likely justified for \lls. We will show later that absorbers with column densities around this value are the dominant contributors to the attenuation length. This explains why the optically thin model gives very similar values for \X\ to the more detailed model of \citetalias{Theuns21}.

The cross section $\sigma$ for which a halo of mass $M_h$ yields a column density higher than a given value of $N_{\rm HI}$ is
\begin{align}
    \sigma(>N_{\rm HI})=\pi b^2(>N_{\rm HI}) = \pi\,R_h^2\times\left(\frac{N_{\rm HI,h}}{N_{\rm HI}}\right)^{2/3}\,.
    \label{eq:sigma}
\end{align}
Some previous models of absorbers \cite[e.g.][]{Fumagalli13, Erkal15} set $\sigma=f_{\rm cov}\,\pi\,R_h^2$, where $f_{\rm cov}$ is a dimensionless \lq covering factor\rq. In our model, Eq.~(\ref{eq:NHh}) shows that {\em even in a spherically symmetric model}, $f_{\rm cov}$ depends on $M_h$, $T$ and $\Gamma_0$, and rather strongly on redshift. We\footnote{The minus sign is, unfortunately, missing in \protect\citetalias{Theuns21}.} now follow \citetalias{Theuns21} by defining the function $g(M_h, N_{\rm HI}, z)$ as the number of absorbers with a given column density per unit co-moving path length $dX$ due to halos of mass $M_h$. \citetalias{Theuns21} shows that this function is proportional to the derivative of the cross-section with respect to $N_{\rm HI}$ times the halo mass function, ${\rm d}n/{\rm d}{\rm log} M_h$,
\begin{align}
g(M_h, N_{\rm HI},z)&\equiv \frac{d^3N}{dN_{\rm HI}\,d\log M_h\,dX}\nonumber\\
&=-\frac{c}{H_0}\frac{dn(M_h, z)}{d\log M_h}\,\frac{d\sigma(M_h, >N_{\rm HI}, z)}{dN_{\rm HI}}\nonumber\\
&=\frac{2c}{3H_0}\,\frac{dn(M_h, z)}{d\log M_h}\,
\frac{\pi R_h^2}{N_{\rm HI,h}}
\,\left(\frac{N_{\rm HI,h}}{N_{\rm HI}}\right)^{5/3}\,.
\label{eq:gCDDF}
\end{align}
The halo mass function, ${\rm d}n/{\rm dlog} M_h$, is the co-moving number density of halos with mass $M_h$ per dex in halo mass. The cross-section
$\sigma$, on the other hand, is defined in {\em proper} units. Therefore the function $g$ will only evolve if the halo mass function evolves in co-moving units, or if the absorbers themselves evolve in proper units, or both.

Integrating the function $g$ over halo mass yields the \cddf,
\begin{align}
f(N_{\rm HI}, z) = \int_{\log M_{\rm crit}(z)}^\infty\, g(M_h, N_{\rm HI}, z)\,d\log M_h\,.
\label{eq:fN}
\end{align}
We note that the lower limit of the integral over halo mass in Eq.~(\ref{eq:fN}) is $\log M_{\rm crit}(z)$, where $M_{\rm crit}(z)$ is the mass below which halos lose their gas when it is photo-heated by the ionizing background. Obviously, such halos will not host absorbers and hence will not contribute to the \cddf. In this paper we use the fit by \cite{Okamoto08} to evaluate $M_{\rm crit}(z)$. It might also be useful to limit the upper limit of integration in Eq.~(\ref{eq:fN}) since gas in sufficiently massive halos is likely to be hot and collisionally ionized, rather than cold and neutral - and hence our model would be a poor description of gas in such halos. Fortunately, such massive halos are rare at the high redshifts $z\ge 2$ that we are mostly interested in, and the steep fall off of the mass function at high $M_h$ implies that such halos contribute negligibly in any case.

Combining all what we found so far allows us to obtain the following analytical expression for the \cddf,
\begin{align}
    f&(N_{\rm HI}, z) = 8.67\times 10^{-19}{\rm cm}^{2}\,
    \frac{f_{17.2}(z)}{\Gamma^{2/3}_{-12}(z)}\,
    \,\left(\frac{N_{\rm HI}}{10^{17.2}{\rm cm}^{-2}}\right)^{-5/3}\nonumber\\
    &\times 
    \left(\frac{f_{\rm gas}}{0.6}\right)^{4/3}\,
    \left(\frac{\alpha_B(T)}{\alpha_B(1.5\times 10^4{\rm K})}\right)^{2/3}\,\,,
\label{eq:fN2}    
\end{align}    
where $f_{17.2}(z)$ is the dimensionless function
\begin{align}    
    f_{17.2}(z) &\equiv \frac{f_c(z)}{f_c(3)}\,\left(\frac{1+z}{4}\right)^{4/3}\,,
    \nonumber\\
  \frac{f_c(z)}{{\rm cMpc}^3} &\equiv \int_{\log M_{\rm crit}(z)}^\infty 
  \frac{dn(M_h, z)}{d\log M_h}\,
\left(\frac{M_h}{10^{10}{\rm M_\odot}}\right)^{8/9}\,d\log M_h\,,\nonumber\\
\label{eq:f17}
\end{align}
and $\Gamma_{-12}\equiv \Gamma_0/(10^{-12}{\rm s}^{-1})$; we note that the normalization $f_{17.2}(z=3)=1$ by construction.

Equation~(\ref{eq:fN2}) brings out the scaling of the \cddf\ with column density, $N_{\rm HI}$, \igm\ temperature, $T$, and photo-ionization rate, $\Gamma_{-12}$ , with any additional redshift dependence encoded by $f_{17.2}(z)$. From now on we will set $f_{\rm gas}=0.6$ and $T=1.5\times 10^4{\rm K}$, and drop them from the equations. If required, the interested reader can always resurrect them by replacing $f_{17.2}\to f_{17.2}\times \,(f_{\rm gas}/0.6)^{4/3}\times \left[\alpha_{\rm B}(T)/\alpha_{\rm B}(T=1.5\times 10^4{\rm K})\right]^{2/3}$.

The explicit redshift dependence of the \cddf\ is encoded by the function $f_{17.2}(z)$, which depends on $f_c(z)$.
The latter dimensionless quantity is approximately\footnote{It would be that fraction if the exponent of $M_h$ in the integral were 1, rather than 8/9.} the mass in a volume of $1~{\rm cMpc}^3$ that is in halos of mass $>M_{\rm crit}(z)$, divided by $10^{10}M_\odot$. This quantity depends on $z$ but is of order unity. The additional redshift dependence for $f_{17.2}$ of $\propto (1+z)^{4/3}$ arises from the $z$ dependence of the relation between halo mass and virial radius. Values of $M_{\rm crit}$ and $f_c$ as a function of redshift are given in Table~\ref{table:modelparams}. To compute the integral over mass, we used the \col\ python package of \cite{Diemer18COLOSSUS}, selecting the implementation of the fit by \cite{Reed07} of the halo mass function, $dn/d\log M_h$.

\begin{table}
\centering
\begin{tabular}{ccc}
\hline
$z$ & $\log M_{\rm crit}$ ($M_\odot$) & $f_c$ \\
\hline
\hline
 0 & 9.82 & 0.94 \\
 1 & 9.57 & 0.98 \\
 2 & 9.35 & 0.87 \\
 3 & 8.99 & 0.77 \\
 4 & 8.71 & 0.65 \\
 5 & 8.42 & 0.55 \\
 6 & 8.19 & 0.46 \\
 \hline
\end{tabular}
\caption{Model parameters as a function of redshift, $z$. $M_{\rm crit}$ is the critical halo mass below which halos lose their baryons, taken from  \protect\cite{Okamoto08}; $f_c$ is the dimensionless variable entering
Eq.~(\ref{eq:f17}).}
\label{table:modelparams}
\end{table}

The analytic optically thin \cddf\ is a power-law in column density, $f\propto N_{\rm HI}^{-5/3}$ (see also \citetalias{Theuns21}); the value of $-5/3$ for the exponent results from the assumed slope of the density profile of gas in halos, $n_{\rm H}(R)\propto R^{-2}$. The model's dependence on $N_{\rm HI}$ agrees well with that of the observed \cddf\, which is also approximately
a power law with slope $-1.66\pm 0.01$ at $\bar z=2.99$ and $-1.68\pm0.02$ at $\bar z=3.48$ at column densities $\lessapprox 10^{16}{\rm cm}^{-2}$ \citep[e.g.][]{Kim21}. \cite{Faucher09} and \cite{Haar12UVbackground} provide more accurate fitting functions for the \cddf\ towards higher and lower $N_{\rm HI}$.

Absorbers with column density $N_{\rm HI}\sim 10^{17.2}{\rm cm}^{-2}$ and higher are particularly important for setting the opacity of the \igm\ to ionising photons, unfortunately, the super-\lls\ range is also where it is very difficult to measure the slope of the \cddf. In addition, it is difficult to provide accurate measurements of the \cddf\ at higher redshifts. Given these observational limitations, it is useful to have an analytical model, such as the one presented here, which predicts the evolution of the \cddf\ and which agrees very well with the data where they are at their most reliable.

The model predicts that over a relatively large range in mass, halos contribute about equally to the \cddf\ per dex in halo mass. The reason for this is at follows. At fixed $N_{\rm HI}$, the cross section $\sigma$ (above which the column density is higher than $N_{\rm HI}$) increases with halo mass $\propto M_h^{8/9}$, with Eq.~(\ref{eq:gCDDF}) elucidating why: $\sigma\propto R_h^2\,N_{\rm HI, h}^{2/3}\propto M_h^{8/9}$. We note, however, that the {\em number density} of halos {\em decreases} with halo mass, approximately $\propto M_h^{-0.9}$ on the power-law part of the \cite{Press74} halo mass function.  As a consequence, all halos with mass above the critical mass, $M_h> M_{\rm crit}$, but below the critical\footnote{As in $dn/d\log M_h\propto M_h^{-\alpha_h}\,\exp(-M_h/M_\ast)$.} Press-Schechter mass $M_\ast$ contribute about equally to the amplitude of the \cddf, with those more massive than $M_\ast$ contributing little.

The redshift evolution of the \cddf\ is a consequence of the following four effects:
({\em i}) the evolution of $\Gamma_{-12}$, ({\em ii}) the evolution of the halo mass function, ({\em iii}) the evolution of $M_{\rm crit}$, and ({\em iv}) the explicit factor\footnote{This factor results from the redshift dependence of the $M_h-R_h$ relation.} $(1+z)^{4/3}$ of Eq.~(\ref{eq:f17}). We examine the impact of the evolution of the \cddf\ on that of the attenuation length in the next section.

\subsection{The evolution of the attenuation length}
\label{sect:L}
\begin{figure*}
\includegraphics[width={0.95\textwidth}]{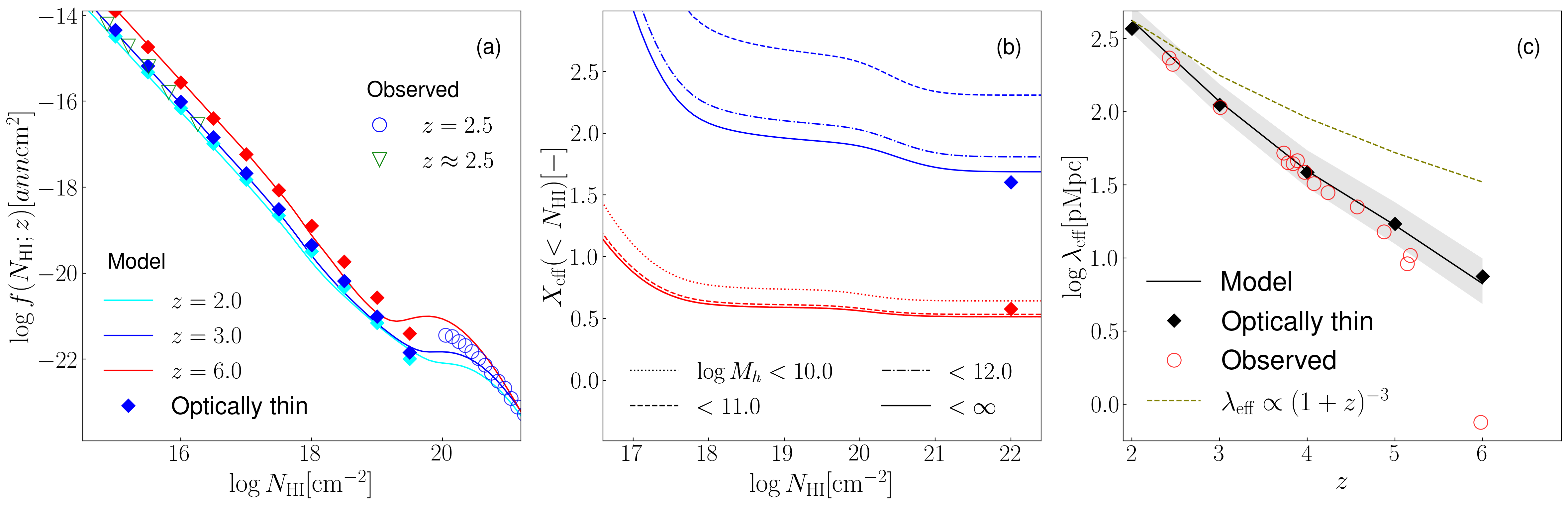}
\caption{Evolution of - {\em left panel}: the \cddf, $f(N_{\rm HI})$, {\em central panel:} attenuation length, \X, plotted cumulatively as a function of column density, {\em right panel:} proper attenuation length \L. {\em Solid lines} are from the model by \protect\cite{Theuns21}, with {\em cyan}, {\em blue} and {\em red solid lines} corresponding to $z=2$, $3$ and $6$, and the {\em black solid line} in the right panel showing the evolution with $z$. {\em Solid diamonds} show the optically thin approximation at those same redshifts, with the \cddf\ in the left panel computed using Eq.~(\ref{eq:fN}), the attenuation length computed using Eq.~(\ref{eq:X}) (central panel), and the corresponding proper attenuation length computed using Eq.(\ref{eq:dX}). In the central panel, the {\em solid curve} includes all halo masses, the {\em dotted} and {\em dashed} and {\em dot-dashed} curves include halos up to $10^{10}$, $10^{11}$, and $10^{12}{\rm M}_\odot$ solar masses. In the right panel, the  {\em shaded area} corresponds to varying $M_{\rm crit}$ (Eq.~\ref{eq:fN}) - the critical halo mass below which halos lose their gas due to photo-evaporation - by a factor of four around the central value taken from \protect\cite{Okamoto08}. Both the optically thin expression (squares) and the results of the model of \protect\citetalias{Theuns21} (solid lines) use the photo-ionization rate $\Gamma_0(z)$ from \protect\cite{Haar12UVbackground}, cosmological parameters from \protect\cite{Planck15} and the critical mass from \protect\cite{Okamoto08}. {\em Coloured empty symbols} are observations: in the {\em left panel}, {\em blue open circles} are the data at $z\sim 2.5$ from \protect\cite{Noterdaeme12}, {\em downward green triangles} are the $z\approx 2.5$ data from \protect\cite{Rudie13}; the {\em open red circles} in the {\em right panel} are the values taken from Fig.~8 of \protect\cite{Becker21} (See text for further details).
}
\label{fig:MF1}
\end{figure*}

\begin{figure*}
\includegraphics[width={0.95\textwidth}]{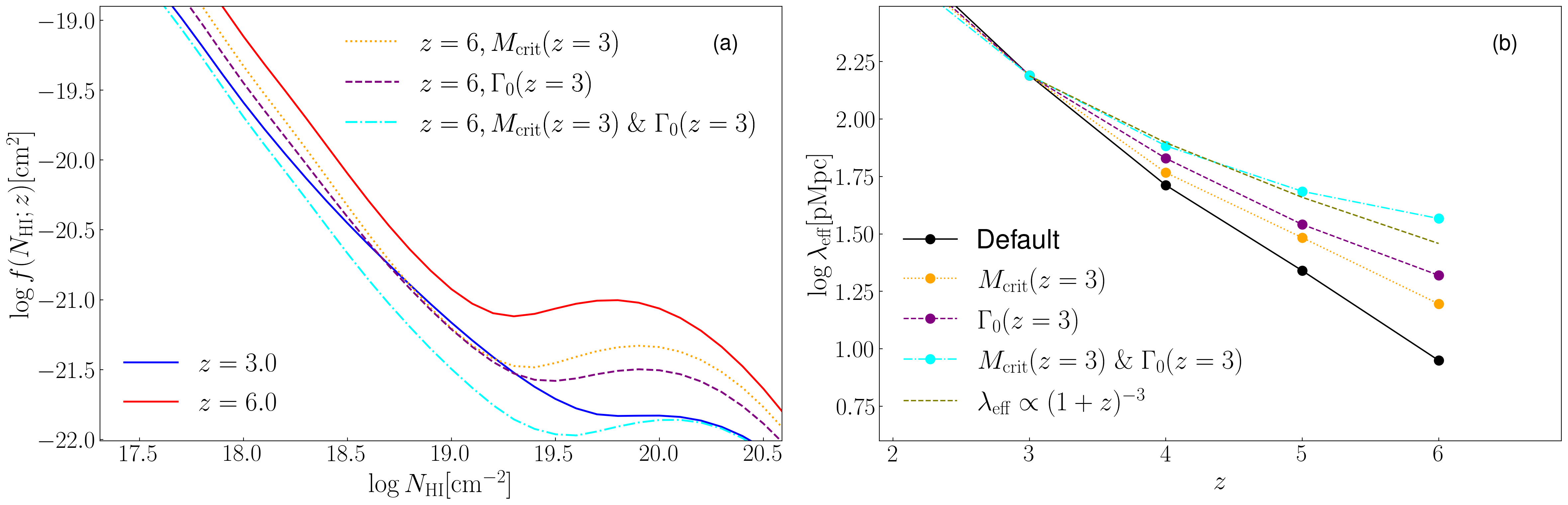}
\caption{Similar to Fig.~\ref{fig:MF1}, with the {\em left panel} showing the evolution of the \cddf, and the right panel the evolution of the proper attenuation length, but this time also illustrating parameter dependencies. The default choice of parameters used in Fig.~\ref{fig:MF1} is shown in {\em blue} and {\em red} in panel (a) for redshift $z=3$ and 6, and black in panel (b). We then computed the \cddf\ at $z=6$ but kept one or more parameters fixed at their value at $z=3$. The {\em orange dotted line}, {\em purple dashed line}, and {\em cyan dot-dashed line} are the \cddf\ at $z=6$, computed using the $z=3$ value of $M_{\rm crit}$, $\Gamma_0$, and both $M_{\rm crit}$ and $\Gamma_0$, respectively. This demonstrates that both parameters affect the evolution of the \cddf\, and when both are kept constant, there is hardly any remaining evolution left. Panel (b) shows the effect of these parameters on the evolution of \L, where $M_{\rm crit}$, $\Gamma_0$, and both $M_{\rm crit}$ are kept fixed at their $z=3$ value for the {\em orange dotted line}, {\em purple dashed line}, and {\em cyan dot-dashed line}. When both parameters are kept constant, \L\ follows closely the evolution $\propto (1+z)^{-3}$, shown as an {\em olive dashed line}. With $M_{\rm crit}$ and $\Gamma_0$ kept constant, the remaining evolution in \L\ is mostly due to cosmological expansion, hence  $\L\ \propto (1+z)^{-3}$.
See the main the text for further discussion.}
\label{fig:MF2}
\end{figure*}

\begin{figure}
\includegraphics[width={0.95\columnwidth}]{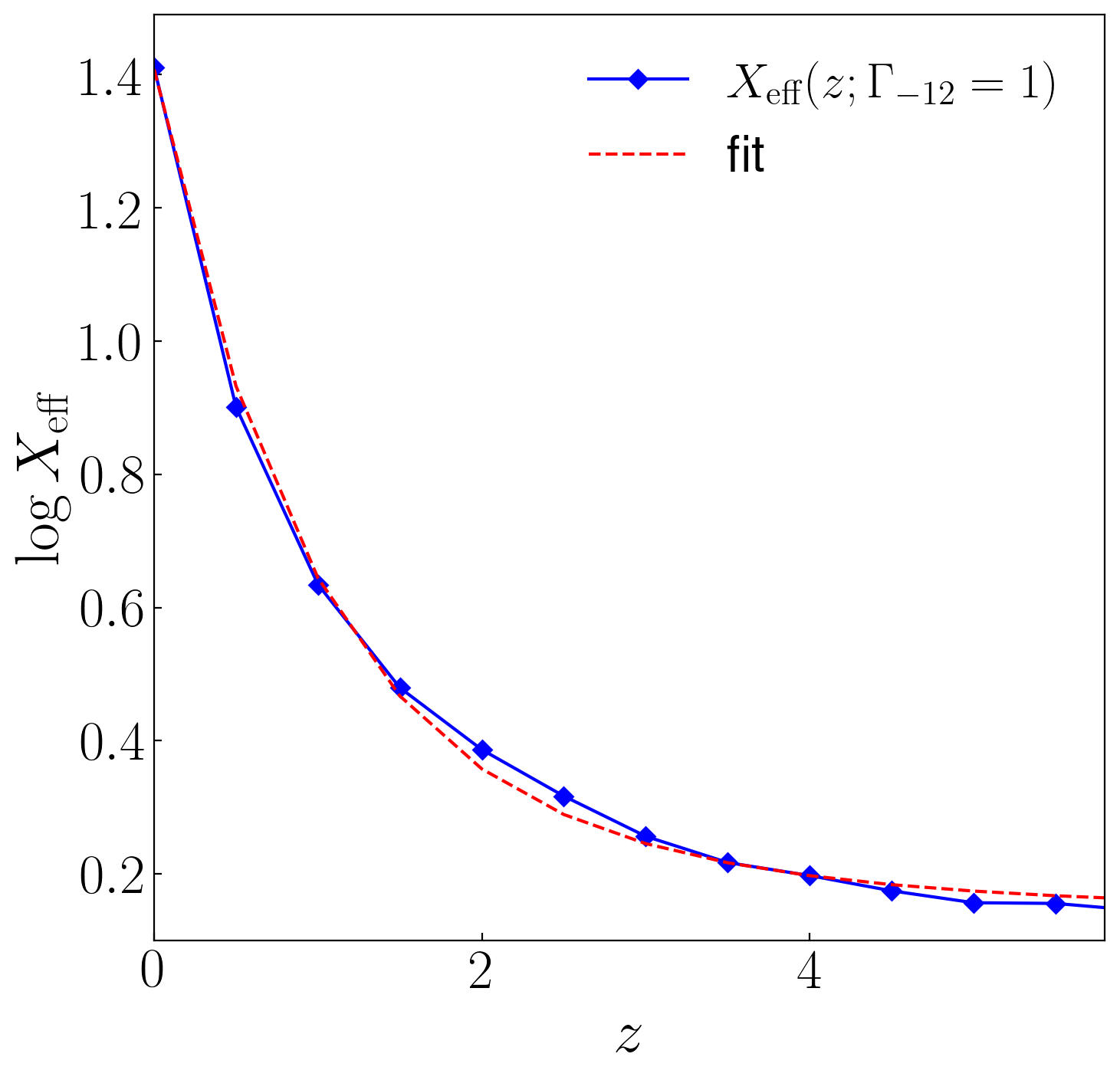}
\caption{Evolution of the co-moving attenuation length, \X, as given by Eq.~(\ref{eq:xeff})
for the case of a constant amplitude of the ionization rate, $\Gamma_{-12}(z)=1$.
The fit shown by the {\em dashed red line} is $\log\X=1.38+24.2/(1+z)^3$, which fits the model
to better than 20 per cent.
}
\label{fig:Xeff}
\end{figure}

We can now combine Eq.~(\ref{eq:X}) for \X\ in terms of the \cddf\ with Eq.~(\ref{eq:f17}) for the shape and evolution of the \cddf. We convert from column density, $N_{\rm HI}$, to optical depth, $\tau$, using $\tau=\sigma_{\rm th}\,N_{\rm HI}$, where $\sigma_{\rm th}$ is the photo-ionization cross section at the Lyman limit ($h\nu_{\rm th}=13.6~{\rm eV}$) and evaluate\footnote{Clearly it is incorrect to integrate from $\tau=0$ to $\tau\to\infty$: 
we have not verified whether the optically thin model reproduces the \cddf\ in the regime of small $\tau$ that corresponds to the Lyman-$\alpha$ forest, and the approximate expression for the \cddf\ is only valid for highly-ionized absorbers and hence not applicable in the regime of {\sc dla}'s. Fortunately, the contribution of very low-$\tau$ absorbers to $X$ is negligible, and we will show that the contribution of high$-\tau$ absorbers depends on their number density but not on $\tau$.} the integral over optical depth between zero and infinity, $\int_0^\infty \tau^{-5/3}\,\left(1-\exp(-\tau)\right)\,d\tau=4.02$. This yields the following expression for the attenuation length\footnote{Where we remind the reader that we have dropped the dependence on $f_{\rm gas}$ and $T$.},
\begin{align}
    \X(z) = &1.80\,
    \frac{\Gamma^{2/3}_{-12}(z)}
    {f_{17.2}(z)}\,\,;
    \label{eq:xeff}
\end{align}
for which the corresponding proper attenuation length is
\begin{align}
\L(z) &= \frac{c\,\X(z)}{H_0\,(1+z)^3}=126\,{\rm pMpc}\,
\frac{\Gamma^{2/3}_{-12}(z)}{f_{17.2}(z)}\,
\left(\frac{4}{1+z}\right)^3\,.
\label{eq:lambda}
\end{align}
The results of our calculations so far are summarised in Fig.~\ref{fig:MF1}. The left panel compares the \cddf\ as computed using the model by \citetalias{Theuns21} (solid lines) to the optically thin approximation of Eq.~(\ref{eq:fN}) at $z=2$ (cyan line and cyan diamonds, respectively), $z=3$ (blue line and blue diamonds)
and $z=6$ (red line and red diamonds). The full model includes self-shielding which causes the transition from $f(N_{\rm HI})\propto N_{\rm HI}^{-5/3}$ in the highly-ionized regime of \lls, to $f(N_{\rm HI})\propto N_{\rm HI}^{-3}$ in the neutral {\sc dla} regime, with the characteristic \lq knee\rq\ between the two power laws around $N_{\rm HI}=10^{20}{\rm cm}^{-2}$ caused by the transition from ionized to neutral absorbers \citep{Zheng02, Erkal15, Theuns21}.
The optically thin model has the same slope and amplitude as the full model in the \lls\ regime. Errors on the observed data are comparable or smaller than the symbols, except for the $z\sim 6$ data point in the right panel which we'll return to later.

The central panel plots the co-moving attenuation length $\X(<N_{\rm HI})$ due to absorbers with columnn density less than $N_{\rm HI}$ in the \citetalias{Theuns21} model, for $z=3$ and $z=6$ (solid lines). The main contribution to \X\ is from absorbers in the relatively small column-density range of $10^{17}{\rm cm}^{-2}\le N_{\rm HI}\le 10^{18}{\rm cm}^{-2}$ at $z=6$, and
$10^{18}{\rm cm}^{-2}\le N_{\rm HI}\le 10^{19}{\rm cm}^{-2}$ at $z=3$. The different line styles show the extent to which halos of a given mass contribute, with dotted, dashed, and dot-dashed lines showing the contribution due to halos with mass less than $10^{10}$, $10^{11}$ and $10^{12}M_\odot$. Halos with mass $> 10^{10}M_\odot$ contribute little to \X\ at $z=6$, but this increases to halos with mass $> 10^{11.5}M_\odot$ by $z=3$. The two diamonds show the value of $\X(z)$ obtained from the optically thin model using Eq.~(\ref{eq:xeff}), with $z=3$ and $z=6$ shown as a blue and a red diamond. Clearly, this approximation captures the results of the more detailed model of \citetalias{Theuns21} very well.

The right panel of Fig.~\ref{fig:MF1} shows the evolution of the proper attenuation length. The solid black line is the evolution computed using the model of \citetalias{Theuns21}. The grey shading shows the effect of varying the value of $M_{\rm crit}$ by factors 1/4 to 4, in order to illustrate how sensitive \L\ is to this parameter. The black diamonds show the optically thin approximation, which captures the evolution of \L\ very well. The solid red circles are the data points plotted in Fig.~8 of \cite{Becker21}.  The data are compiled from \cite{Proc09opacity, Omeara13, Fumagalli13, Worseck14} and \cite{Lusso18}, with the highest $z$ point from \cite{Becker21}.

The model reproduces the observations well over the range $z=2\to 5$, and this is one of the main results of this paper. As a note of caution, we note that the value taken for $f_{\rm gas}$ affects \L, yet our choice of taking $f_{\rm gas}=0.6$ is not particularly well motivated. We suspect that this parameter attempts to account for the fact that the neutral gas distribution in real absorbers is not spherically symmetric \citep[see {\em e.g.} the analysis by ][]{Erkal15}. \citetalias{Theuns21} use $f_{\rm gas}=0.5$ (rather than 0.6) in their model for \dla's, and hence a value of $f_{\rm gas}\sim 0.6$ fits the \cddf\ all the way from \lls\ to \dla's at $z=3$, as can be seen in the left panel of the figure. Strikingly, the model does not show the dramatic decline in \L\ suggested by the data from \cite{Becker21} from $z=5\to 6$: we will return to this in section \ref{sect:bias}.

The gold-dashed line in the right panel of Fig.~\ref{fig:MF1} shows the scaling $\propto (1+z)^{-3}$. Both data and model evolve faster than this, implying that the absorbers either evolve in co-moving number density or proper size, or both. We examine the cause of the enhanced evolution in the model in more detail in Fig.~\ref{fig:MF2} as follows: we redo the calculations but we keep the value of $M_{\rm crit}(z)$ and $\Gamma_{-12}(z)$ constant and equal to their values at $z=3$: this is the cyan curve in both panels. The left panel shows that in this case, the $z=6$ \cddf\ is almost identical to the $z=3$ \cddf. We note that the main remaining difference is the evolution of the halo mass function, but that clearly has relatively little effect on the \cddf. The reason is that the halo mass function evolves relatively little below $M_\ast$, and halos above $M_\ast$ where the halo function does evolve rapidly contribute little to \L. 

The right panel of Fig.~\ref{fig:MF2} shows the effect of $M_{\rm crit}$ and $\Gamma_{-12}$ separately. Both the evolution of $M_{\rm crit}$ and of $\Gamma_{-12}$ contribute\footnote{See also \cite{Cain23}.} to the evolution of \X, causing $\X(z)$ to increase with decreasing $z$. When these parameters are kept constant, \X\ evolves much less, and \L\ evolves mostly due to cosmological expansion, $\L\propto (1+z)^{-3}$. This can be seen by the fact that the cyan line - for which $M_{\rm crit}$ and of $\Gamma_{-12}$ both remain constant - falls almost on top of the $\L\propto (1+z)^{-3}$ scaling. The right panel also shows that $M_{\rm crit}$ and $\Gamma_{-12}$ contribute about equally to the evolution of \X. We plot the evolution of $X_{\rm eff}(z)$ for the case of a constant amplitude of the ionization rate ($\Gamma_{-12}(z)=1$) in Fig.~\ref{fig:Xeff}. 

Summarizing: the attenuation length \X\ evolves due to the evolution of $M_{\rm crit}$ and $\Gamma_{-12}$. $M_{\rm crit}$ is the critical mass below which halos lose or cannot accrete gas. The evolution in $M_{\rm crit}$ is itself mostly caused by the fact that halos cannot accrete gas if their virial temperature is lower than the temperature of the gas they attempt to accrete - and the virial temperature of a halo of given mass depends on $z$ - hence the evolution. At lower $z$, $M_{\rm crit}$ is higher, and so a larger fraction of halos no longer host the absorbers that limit \X, and hence \X\ increases. About equally important to the evolution of \X\ is that $\Gamma_{-12}(z)$ increases with decreasing $z$ (from $z=6\to 2$), making the gas in the absorbers more highly ionized, which again increases \X. Finally, we note that the co-moving number density of absorbers is proportional to the halo mass function, which, of course, {\em increases} with decreasing $z$. So, despite that the number density of absorber hosts increases, the net absorption they produce decreases and hence \X\ increases with cosmic time. We recall that the number density of halos on the power-law tail of the Press-Schechter mass function does not actually evolve strongly.

The values of \X\ and \L\ in the current model are set to a large extent by the number density of absorbers with $N_{\rm HI}\sim 10^{18}{\rm cm}^{-2}$. It is possible to directly count the number density of such strong absorbers in {\sc qso} spectra.
We compare these predictions to observations in Appendix~\ref{sect:lls}. Because absorbers are associated with halos in the current model, it is straightforward to infer the clustering of absorbers from the clustering of their host halos. This is what we'll do next.

\section{Clustering of absorbers}
\label{sect:bias}
\begin{figure*}
\includegraphics[width={0.95\textwidth}]{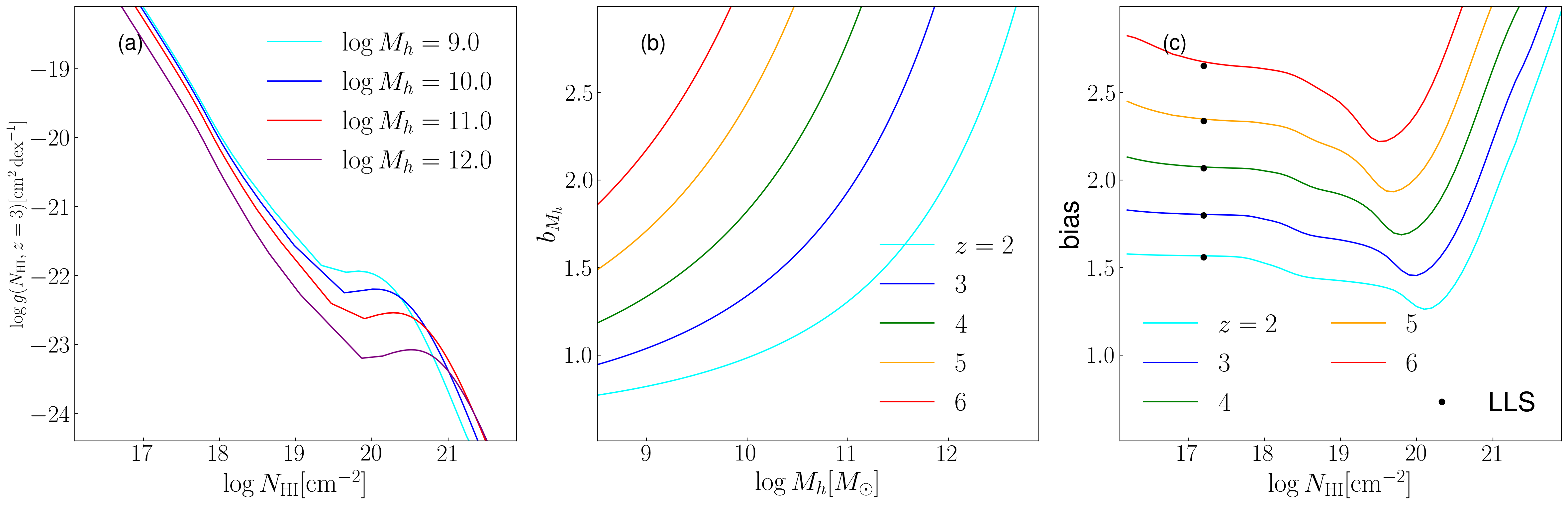}
\caption{Bias of absorption systems in the model of \protect\cite{Theuns21}, as computed
from Eq.~(\ref{eq:bias1}). {\em Panel (a):} contribution of halos of a given mass to the \cddf\ at redshift $z=3$, as given by Eq.~(\ref{eq:gCDDF}). The somewhat artificial shape around $N_{\rm HI}\sim 10^{19}{\rm cm}^{-2}$ is due to our interpolation of the temperature of the gas from $T=15000~{\rm K}$ when optically thin to $T=10^4~{\rm K}$ for $\tau\ge 1$. {\em Panel (b):} bias of halos as a function of their mass, $b_{\rm M_h}$, at different redshifts, as computed with \col\ \protect\citep{Diemer18COLOSSUS}. {Panel (c):} Bias for lines of a given column density, $b_{\rm N_{\rm HI}}$ from Eq.~(\ref{eq:bias1}), for different redshifts; {\em black circles} show the bias for Lyman-limit systems, $b_{\rm LLs}$ from Eq.~(\ref{eq:biasLLS}), for those same redshifts; these points are plotted at $N_{\rm HI}=10^{17.2}{\rm cm}^{-2}$. The bias of \lls's (absorbers with $N_{\rm HI}\geq 10^{17.2}{\rm cm}^{-2}$) is very close to that of lines with a column density of $10^{17.2}{\rm cm}^{-2}$. Coloured lines in panel (a) correspond to different halo masses, and in panels  (b) and (c) correspond to different redshifts, as per the legends. See text for discussion.}
\label{fig:Bias}
\end{figure*}

In this section we compute the bias ($b$) of absorbers as a function of their column density, finding that $b$ is nearly independent of $N_{\rm HI}$. Somewhat surprisingly, we find that the bias of \dla's with $N_{\rm HI}\sim 10^{20.3}{\rm cm}^{-2}$ is actually {\em lower} than that of \lls's and sub-\lls's. We use the
bias-$N_{\rm HI}$ relation to investigate the impact of bias on \L\ in \S~\ref{sect:b_L}, showing that the bias of quasars likely impacts the measured values of \L\ significantly above $z\sim 5$. In \S~\ref{sect:biaslambda}, we use these findings to compute the probability distribution of $\tau_{\rm eff}$.

\subsection{The bias of absorbers as a function of $N_{\rm HI}$}
\label{sect:halobias}
The bias of absorbers with a given hydrogen column density at redshift $z$ follows from that of their host halos as (\citetalias{Theuns21})
\begin{equation}
b_{N_{\rm HI}}(z) = 
\frac{
\int_{\log M_{\rm crit}(z)}^{\infty}
d\log M_h\,\left\{b_{M_h}(z)\times g(N_{\rm HI}, M_h, z)\right\}
}
{
\int_{\log M_{\rm crit}(z)}^{\infty}
d\log M_h\,\left\{g(N_{\rm HI}, M_h, z)\right\}
}\,,
\label{eq:bias1}
\end{equation}
and the bias of absorbers with $N_{\rm HI}\geq 10^{17.2}{\rm cm}^{-2}$ is
\begin{eqnarray}
&\hphantom{.}&\hspace{-1cm}b_{\rm LLs}(z)=\nonumber\\
&\hphantom{..}&\hspace{-2cm}
\frac{
\int_{17.2}^{\infty} d\log N_{\rm HI}\int_{\log M_{\rm crit}(z)}^\infty d\log M_h\,
{\cal F}_1(M_h, N_{\rm HI}, z)
}
{ 
\int_{17.2}^{\infty} d\log N_{\rm HI}\int_{\log M_{\rm crit}(z)}^\infty d\log M_h\,
{\cal F}_2(M_h, N_{\rm HI}, z)
}\nonumber\\
{\cal F}_1(M_h, N_{\rm HI}, z) &=& b_{M_h}(z)\,N_{\rm HI}\, g(N_{\rm HI}, M_h, z)\nonumber\\
{\cal F}_2(M_h, N_{\rm HI}, z) &=& N_{\rm HI}\, g(N_{\rm HI}, M_h, z)\,.
\label{eq:biasLLS}
\end{eqnarray}
Here, $b_{M_h}(z)$ is the bias of a halo of virial mass $M_h$ at redshift $z$, and column densities are assumed to be expressed in units of cm$^{-2}$. 

The bias computed from Eq.~(\ref{eq:bias1}) for absorbers with a given column density is plotted in panel (c) of Fig.~\ref{fig:Bias}, with colours indicating redshift. Below column densities of $\sim 10^{18.5}{\rm cm}^{-2}$, absorber bias is nearly independent of column density. At first somewhat surprising, we also find that the bias then {\em decreases} with increasing column density, until it reaches a minimum value for $N_{\rm HI}\sim 10^{20.3}{\rm cm}^{-2}$, after which 
the bias increases rapidly with increasing column density.

These trends can be understood by examining panel (a) of Fig.~\ref{fig:Bias}, where we plot the function $g(M_h, N_{\rm HI},z )$ defined in Eq.~(\ref{eq:gCDDF}) at 
a representative redshift\footnote{The trends with halo mass are similar at other redshifts.} $z=3$. Below a column density of $\sim 10^{18.5}{\rm cm}^{-2}$, the relative contribution of halos as a function of mass varies little with column density because $g\propto N_{\rm HI}^{-5/3}$, independently of halo mass.
Since all halos contribute about equally to the number density of lines with a given $N_{\rm HI}$, it follows that the bias is independent of $N_{\rm HI}$.

However, the nature of absorbers changes from mostly ionized to mostly neutral at higher column densities, $N_{\rm HI}\sim 10^{19}{\rm cm}^{-2}$.  This transition imprints the \rq knee\rq-shaped feature in $g$ and also in the \cddf\ \citep{Zheng02Lya, Erkal15, Theuns21}. Lower mass halos transition from ionized to neutral at lower values of $N_{\rm HI}$ compared to higher mass halos, as can be seen in panel (a) of Fig.\ref{fig:Bias}. This results in a decrease in the absorber bias because more of these absorbers are associated with lower mass halos - which themselves are less biased. Once $N_{\rm HI}$ is high enough so that most absorbers  have made the transition from highly ionized to neutral, the bias increases rapidly with increasing $N_{\rm HI}$. This is because, in this regime, higher column densities are increasingly associated with more massive - and hence more highly biased - halos - as is also apparent from panel (a) of Fig.\ref{fig:Bias}.

The numerical value of the bias and its evolution with redshift can be understood by also examining panel (b) in Fig.~\ref{fig:Bias}, where we plot the halo bias, $b_{M_h}(z)$, computed using \col\ \citep{Diemer18COLOSSUS}. Panel (a) shows that halos in the mass range $10^9-10^{11}M_\odot$ contribute about equally to the \cddf\ at column densities $N_{\rm HI}\leqslant 10^{18.5}{\rm cm}^{-2}$ at $z=3$, resulting in a weighted bias of such absorbers of $\sim 1.8$, a bit less than that of halos of mass $10^{11}M_\odot$. With increasing redshift, the contribution of lower mass halos increases compared to that of more massive halos at a given value of $N_{\rm HI}$ - which would lower the bias. However, the bias of these same lower mass halos {\em increases} rapidly with redshift. The net result of these opposing trends is an increase in the bias of absorbers with increasing $z$, as seen in panel (c).

Current measurements of the \dla\ bias yield values that range from $b=1\to 3$ \citep[e.g.][]{Alonso18, Perez18, Perez23}. Given the strong dependence of $b$ on $\log N_{\rm HI}$ and redshift in the model, a fair comparison between model and data requires careful modelling of the observational selection which we have not performed yet.

Finally, panel (c) also shows the bias of \lls's computed using Eq.~(\ref{eq:biasLLS}) as filled black dots. With the function $g$ decreasing rapidly with increasing $N_{\rm HI}$, it is not surprising that the bias of \lls's is close to that of absorbers with $N_{\rm HI}=10^{17.2}{\rm cm }^{-2}$. In the next section, we use the bias of absorbers to calculate the attenuation length of photons that themselves emanate from a biased region.

\subsection{The impact of bias on the attenuation length}
\label{sect:b_L}
\begin{figure}
\includegraphics[width={0.95\columnwidth}]{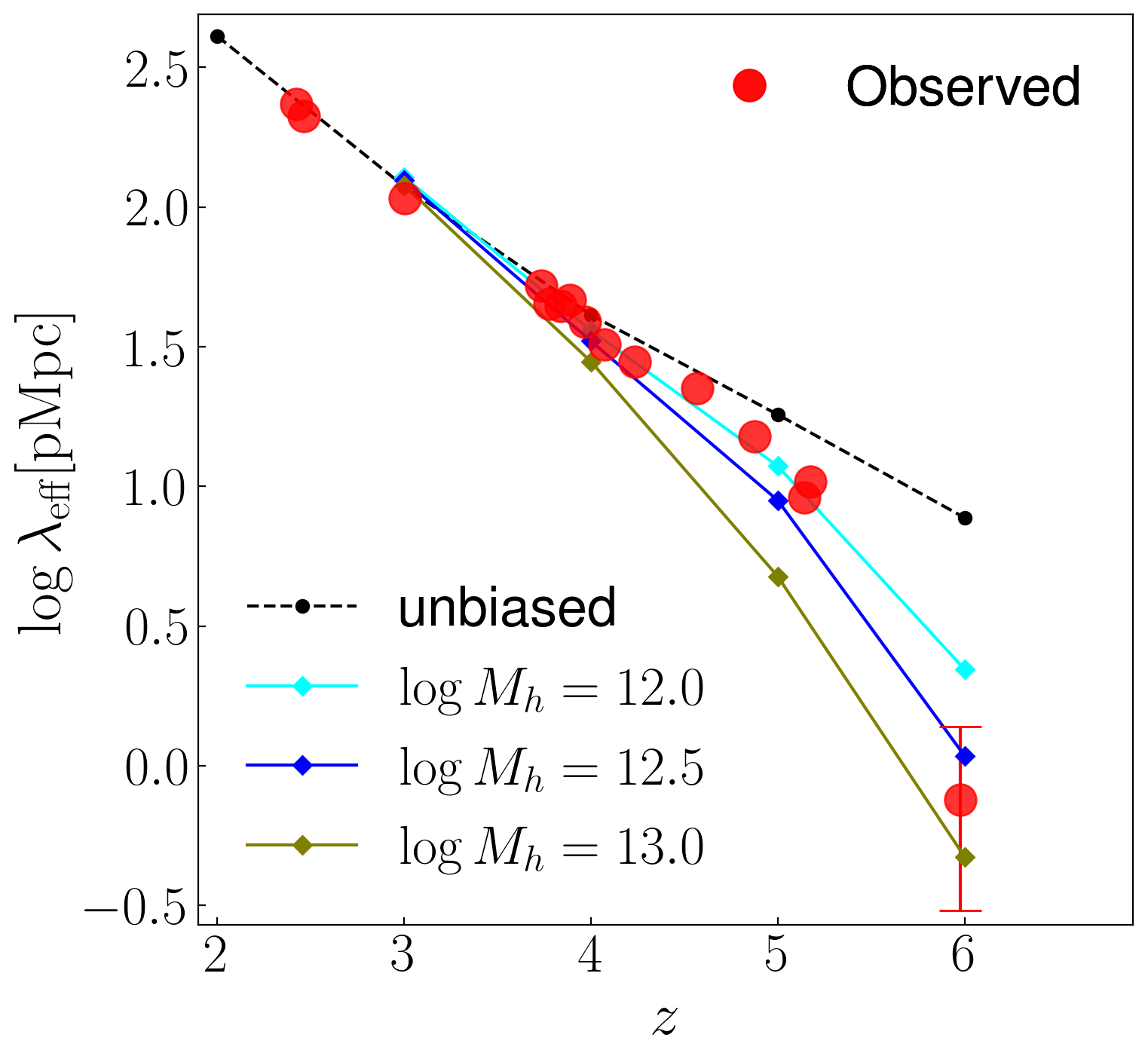}
\caption{Proper attenuation length \L\ as a function of redshift, $z$. The 
{\em dashed black line} is the unbiased model repeated from Fig.~\ref{fig:MF1} (where it was labelled \lq model\rq). The other curves include the effects of the absorber and source bias from Eq.~(\ref{eq:Xb}). The bias of Lyman-limit systems at redshift $z$ is computed as in section \ref{sect:halobias}; the bias of the host halo of the \qso\ is computed for various halo masses $M_h$ as indicated in the legend. {\em Red symbols} repeat the observational data from Fig.~\ref{fig:MF1}, with the $z=6$ data point additionally displaying the uncertainty in the measured value of \L\ taken from \protect{\citealt{Becker21}}.
}
\label{fig:MFP-bias}
\end{figure}

Up to now, we calculated the attenuation length \L\ in the general \igm. However, in observations, \L\ is measured from \qso\ spectra. Given that \qso's may well predominantly be hosted by massive halos that are biased, particularly at higher redshifts, {\em observed} values of \L\ are potentially  biased. We can use the absorber bias determined in the previous section to examine the importance of both sources of bias (\qso\ and absorber) on the measured value of \L\ as follows.

Consider absorbers with column density $N_{\rm HI}$, located at a proper distance between $l$ and $l+dl$ from a source (typically a \qso). The contribution of such absorbers to the effective optical depth at the Lyman limit towards that source is on average
\begin{align}
    \langle d\tau_{\rm eff}\rangle &= \langle N\rangle\,[1-\exp(-\tau)]\nonumber\\
    \langle N\rangle &= f(N_{\rm HI})\,\,dN_{\rm HI}\,\frac{dX}{dl}{dl}\,,
\end{align}
according to Eq.~(\ref{eq:dtaudl}), where $\langle N\rangle$ is the average number of these absorbers and 
$\tau=\sigma_{\rm th}N_{\rm HI}$ is the optical depth of a single absorber at the Lyman limit; $dX/dl$ is given by Eq.~(\ref{eq:dX}). In the absence of clustering, $f(N_{\rm HI})$ is independent of $l$, and hence so is the average number of absorbers, $\langle N\rangle$.

With bias of both absorbers and source accounted for, $\langle N\rangle$ changes to\footnote{{This assumes the linear halo bias model of \cite{Mo10}.}}
\begin{align}
    \langle N\rangle &= f(N_{\rm HI})\,\left\{1+b_S\,b_{N_{\rm HI}}\,\xi(l)\right\}\,dN_{\rm HI}\,\frac{dX}{dl}dl\,.
\end{align}
The factor $\left\{1+b_S\,b_{N_{\rm HI}}\,\xi(l)\right\}$ accounts for linear bias between absorbers and source; $\xi(l)$ is the correlation function of the mass. In the parlance of halo bias, we note that this accounts for the \lq two-halo\rq\ term, {\em i.e.} the clustering of the halos hosting absorber and source, rather than the fact that the host halo of the source may itself host an \lq associated\rq\ absorber (which would be the \lq one-halo\rq\ term, due to absorbers within the host galaxy of the \qso, associated with its own circum-galactic medium, its satellite galaxies or with Magellanic Stream-like features, say).  

We now take advantage of the findings in the previous section that the bias of an absorber, $b_{N_{\rm HI}}$, is approximately independent of column density for the column densities below $10^{19}{\rm cm}^{-2}$ that dominate the attenuation. Therefore, it is a good approximation to replace $b_{N_{\rm HI}}\to b_{\rm LLS}$. We can now compute the attenuation length when accounting for bias, \Xb, in terms of its unbiased value, \X, by using Eq.~(\ref{eq:X}),
\begin{align}
    \int_0^{\Xb}\, \left[1+b_Sb_{\rm LLS}\xi(X)\right]\,dX = \X\,.
    \label{eq:Xb}
\end{align}

Since $b_{\rm LLS}>1$ and $b_S$ and $\xi$ can be significantly larger than 1, $\Xb<\X$: since there are (possibly many) more absorbers close to the \qso\ per unit $dX$ than in the general \igm, the attenuation length measured in the spectra of a \qso\ is generally shorter than its value in the general \igm.

An easy way to account for biasing is to define the dimensionless variable $dY$ by
\begin{align}
    dY &\equiv \left[1+b_S\,b_{\rm LLS}\,\xi(X)\right]\,dX\,,
    \label{eq:Y}
\end{align}
with boundary condition\footnote{$\xi$ is usually expressed as a function of co-moving distance, $l(1+z)$, but we find it more convenient to express $\xi$ as a function of $X$.} $Y=0$ for $X=0$. The statistical properties of the effective optical depth out to $X$ depends on $Y$, which we dub \lq biased absorption length\rq. Using $Y$, rather than $X$, allows us to include the effects of the clustering of absorbers with sources of ionizing photons easily. The average number of absorbers that contribute to $d\tau_{\rm eff}$ in a narrow interval of biased absorption length $[Y, Y+dY]$ from a source is then simply
\begin{equation}
    \langle N(Y)\rangle=f(N_{\rm HI}) dN_{\rm HI}\,dY\,,
\end{equation}
and looks identical to the case where bias is neglected - except for the change of variables $X\to Y$.

We used the \col\ {\sc python} package of \cite{Diemer18COLOSSUS} to compute the correlation function $\xi$ at several redshifts, and then evaluated Eq.~(\ref{eq:Xb}) to compute \Xb\ and \Lb, the attenuation lengths at the the Lyman limit when accounting for bias; the results are shown in Fig.~\ref{fig:MFP-bias}. As illustrative examples, we plot \Lb\ 
when the bias of the source equals that of halos of mass $10^{12}$, $10^{12.5}$, and $10^{13}{\rm M}_\odot$ (cyan, dark blue and olive line) with the bias of the absorbers as calculated in the previous section. In contrast, the unbiased case repeated from Fig.~\ref{fig:MF1} is shown as a dashed black line. \Lb\ decreases with increasing source bias (increasing halo mass), as expected. Around $z\sim 3$, even halos of mass $10^{12.5}M_\odot$ are not that strongly biased to make \Lb\ differ significantly from \L. However, $b_S$ for such halos increases rapidly with increasing $z$ (see e.g. the middle panel of Fig.~\ref{fig:Bias}), and at $z=6$, $\Lb\ll \L$. \qso\ host halo masses of $\sim 10^{12.5}M_\odot$ (dark blue line) bring the computed value of the attenuation length in better agreement with the data, and also reproduces the rapid decreases in the measured value of \Lb\ towards $z=6$. Values of $M_h\sim 10^{12.5}{\rm M}_\odot$ are expected for the host halo masses of $z\sim 6$ \qso's (see \citealt{Zhang23, Beers23} and references therein, see also \citealt{Bower17} for a more general physical model for what sets the halo mass of bright {\sc agn}).

{The sudden decrease in \Lb\ from $z=5\to 6$ in our model is due to the rapid increase in bias of the host halo of the \qso\ in which \Lb\ is measured (itself a consequence of the host halo mass being on the exponential part of halo mass function). Several recent papers instead investigate the possibility that this drop is because this redshift range probes the tail-end of reionization \citep[e.g.][]{Keating20, Daloisio20, Cain21,  Garaldi22, Gaikwad23}. If this were correct, the drop might be due to a rapid change in the emissivity of ionizing photons and/or in the clumping factor of the {\sc igm}. Which interpretation is correct? We first note that the number density of \qso's with 1450\AA\ magnitude brighter than -26 ({\em i.e.} comparable to those of the XQR-30 sample presented by \citealt{Bosman22}
and used by \citealt{Gaikwad23}) - is $\sim 10^{-9}\,{\rm cMpc}^{-3}{\rm mag}^{-1}$ at $z\sim 6$ \citep{Onoue17}.
This implies that even the largest simulation volume investigated
in these papers (of order 160~cMpc$^3$) contains on average only $\sim 4\times 10^{-3}$ \qso's as luminous (and hence plausibly as biased) as those observed. This illustrates the challenge of performing simulations that resolve the physically small absorbers in a simulation that is large enough to also contain the kind of background sources against which we detect them observationally. It also means that these papers cannot test the impact of bias discussed in this paper. Obviously, even if bias plays an important role, it is still possible that this redshift range probes the tail-end of reionization: the two explanations are not mutually exclusive.}

\subsection{The \pdf\ of the biased attenuation length}
\label{sect:biaslambda}
\begin{figure*}
\includegraphics[width={0.95\textwidth}]{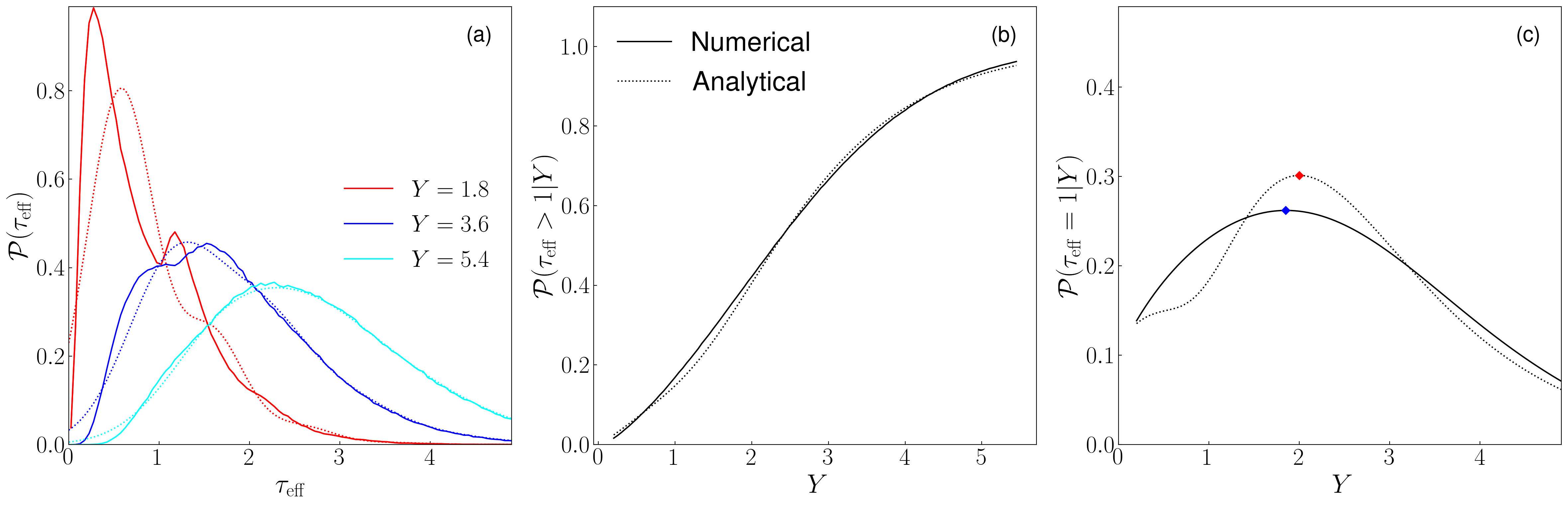}
\caption{Statistics of the effective optical depth, $\tau_{\rm eff}$, and of the co-moving attenuation length, in terms of the (biased) absorption path length $Y$ defined in Eq.~(\ref{eq:Y}). {\em Solid lines} are numerical results, obtained by generating Poisson-distributed absorbers numerically; {\em dashed lines} use the analytical approximation described in the text. {\em Left panel:} Probability distribution of $\tau_{\rm eff}$ for the values $Y$ indicated in the legend. The analytical expression is Eq.~(\ref{eq:taueff}). {\em Central panel:} fraction of paths that reach $\tau_{\rm eff}>1$ within a length $Y$. The analytical expression is Eq.~(\ref{eq:taugt1}). {\em Right panel:} probability that a path with length $Y$ reaches $\tau_{\rm eff}>1$; the mean attenuation length is $Y_{\rm eff}=1.8$. See text for further discussion.
}
\label{fig:MFP}
\end{figure*}
We calculated the mean value of the biased attenuation length in the spectrum of a \qso\ in the previous section as an integral of $f(N_{\rm HI})$, where the \cddf\ is the {\em mean} number of absorbers with a given column density $N_{\rm HI}$ per $dX$. However, a given sight line may have slightly more or slightly fewer lines than that  mean number. As a consequence, the effective optical depth of a given sight line with a given extent $\Delta X$ may be larger or smaller than the ensemble average. To quantify this, we compute in this section  ${\cal P}(\tau_{\rm eff}| X)$ - the {\sc pdf} of the effective optical depth for a sight line with a given co-moving path length $X$. Similarly, we defined and computed the co-moving attenuation length as that value of $X$ for which $\tau_{\rm eff}=1$. Accounting for variations in the number of absorbers along different sight lines, we can compute ${\cal P}(\tau_{\rm eff}=1, X)$ - the probability that $\tau_{\rm eff}=1$ for a given absorption path length. These {\sc pdf}'s may be useful when interpreting observations that are based on a relatively small number of independent sight lines. It is straightforward to account for bias in these calculations by using $dY$ rather than $dX$, but we think that our analysis is easier to follow when we perform the calculation in terms of $dX$.

We will assume that the absorbers are {\em Poisson-distributed}, so that the probability ${\cal P}(N)$ for finding $N$ absorbers in a region where the mean number is $\langle N\rangle$ is given by
\begin{align}
    {\cal P}(N) &= \mathbb{P}(N|\langle N\rangle) \equiv \frac{\langle N\rangle^{N}\,
    \exp(-\langle N\rangle)}{N!}\,,
\end{align}
where $\mathbb{P}(n|\mu)$ is the Poisson distribution with mean $\mu$. In terms of the contribution of such absorbers to the effective optical depth, the \pdf\ of $d\tau_{\rm eff}$ follows from that of $N$ by a change of variables,
\begin{equation}
    {\cal P}(d\tau_{\rm eff}) = \mathbb{P}(N| \langle N\rangle)\,\frac{1}{1-\exp(-\tau)}\,,
\end{equation}
with mean $\langle N\rangle\,(1-\exp(-\tau))$ and dispersion 
$\langle N\rangle\,(1-\exp(-\tau))^2$. 

The total effective optical depth is obtained by integrating $d\tau_{\rm eff}$ over all column densities, but there is no simple relation between the Poisson statistics of the lines and the \pdf\ of $\tau_{\rm eff}$. This is because a linear combination of Poisson distributed variables is not Poisson distributed (or indeed has any other simple \pdf\footnote{
See {\em e.g.} \cite{Bohm14} for a discussion of such \lq Compound Poisson distributions\rq.}).
We can generate Poisson-distributed variables for all $N$'s ({\em i.e.} absorbers with a given small range in column density) and sum $\tau_{\rm eff}$ in bins of $dN_{\rm HI}$ and $dX$, and compute the \pdf\ of $\tau_{\rm eff}$ numerically. It is also possible to derive an approximate analytical expression for the {\sc pdf}. The approximation consists of assuming that absorbers with column below some value (we use $10^{17.2}{\rm cm}^{-2}$) are sufficiently numerous that we can apply the central limit theorem and take them to be Gaussian distributed. This allows the calculation of the {\sc pdf} for \lq low\rq\ $\tau$ absorbers. The stronger absorbers then all have transmission $\exp(-\tau)\approx 0$, and we can then also calculate their {\sc pdf}. Summing the contribution of low and high $\tau$ absorbers yields the net {\sc pdf}, see  Appendix~\ref{sect:AppendixA} for full details.

The results are illustrated in Fig.~\ref{fig:MFP}, where we plot them in terms of $Y$ rather than $X$, with the change of variables accounting for the bias of absorbers and source. For illustrative purposes we assume a \cddf\ of the form of Eq.~(\ref{eq:fN2}), 
\begin{align}
    f(N_{\rm HI}) = 1.27\times 10^{-18}{\rm cm}^{2}\,\left(\frac{10^{17.2}{\rm cm}^{-2}}{N_{\rm HI}}\right)^{5/3}\,,
\end{align}
for which $Y_{\rm eff}=1.8$. We draw Poisson distributed absorption lines from this \cddf\ in narrow bins of $N_{\rm HI}$, which allow us to compute $\tau_{\rm eff}$ for a given biased absorption distance $Y$. We can use this
to compute the fraction of paths that reach $\tau_{\rm eff}>1$ within a given value of $Y$, and the fraction of paths that reach $\tau_{\rm eff}>1$ in a narrow interval $dY$ around $Y$. These are shown as solid lines in panels (a)$\to$(c). The corresponding analytical expressions, Eqs.~(\ref{eq:taueff}), (\ref{eq:taugt1}) and (\ref{eq:pdfY}) derived in Appendix~\ref{sect:AppendixA} are plotted with dashed lines.

When $Y$ is small - the case $Y=1.8$ (which is equal to the biased attenuation length) in panel (a) -  the \pdf\ of $\tau_{\rm eff}$ has two clear maxima, which correspond to 0 or 1 strong absorbers contributing to $\tau_{\rm eff}$ (there are further oscillations visible, due to 2 or more strong absorbers). Absorbers with $\tau\lessapprox 10^{17.2}{\rm cm}^{-2}$ are sufficiently rare when $Y$ is small that approximating their \pdf\ as Gaussian is not very accurate. This is the reason that the analytical model differs noticeably from the numerical calculation for small values of $Y$. As $Y$ increases, such absorbers become more common and the approximation improves.

The analytical model reproduces rather well the fraction of paths that reach $\tau_{\rm eff}>1$ within a given path length $Y$, as shown in panel (b). The derivative of this function with respect to $Y$ is the probability that a given path reaches $\tau_{\rm eff}>1$ in a small interval $dY$ around $Y$, and is plotted in panel (c). The peak of the analytical approximation (red dot) is a bit narrower than that of the numerical result (blue dot), but the location of the maxima is very close. Both functions have a long tail to large values of $Y$, with the analytical approximation reproducing the numerical result well.

Note that the attenuation length in the case shown is $Y_{\rm eff}=1.8$ - yet less than 30 per cent of sight lines with path length $Y=Y_{\rm eff}$ reach $\tau_{\rm eff}=1$ because the distribution of ${\cal P}(\tau_{\rm eff}=1, Y)$ around the mean is quite wide. This is of course because absorption is dominated by the rare, high column density absorbers.
\section{A direct measure of the attenuation length}
\label{sect:measuring}

\begin{figure}
\includegraphics[width={0.95\columnwidth}]{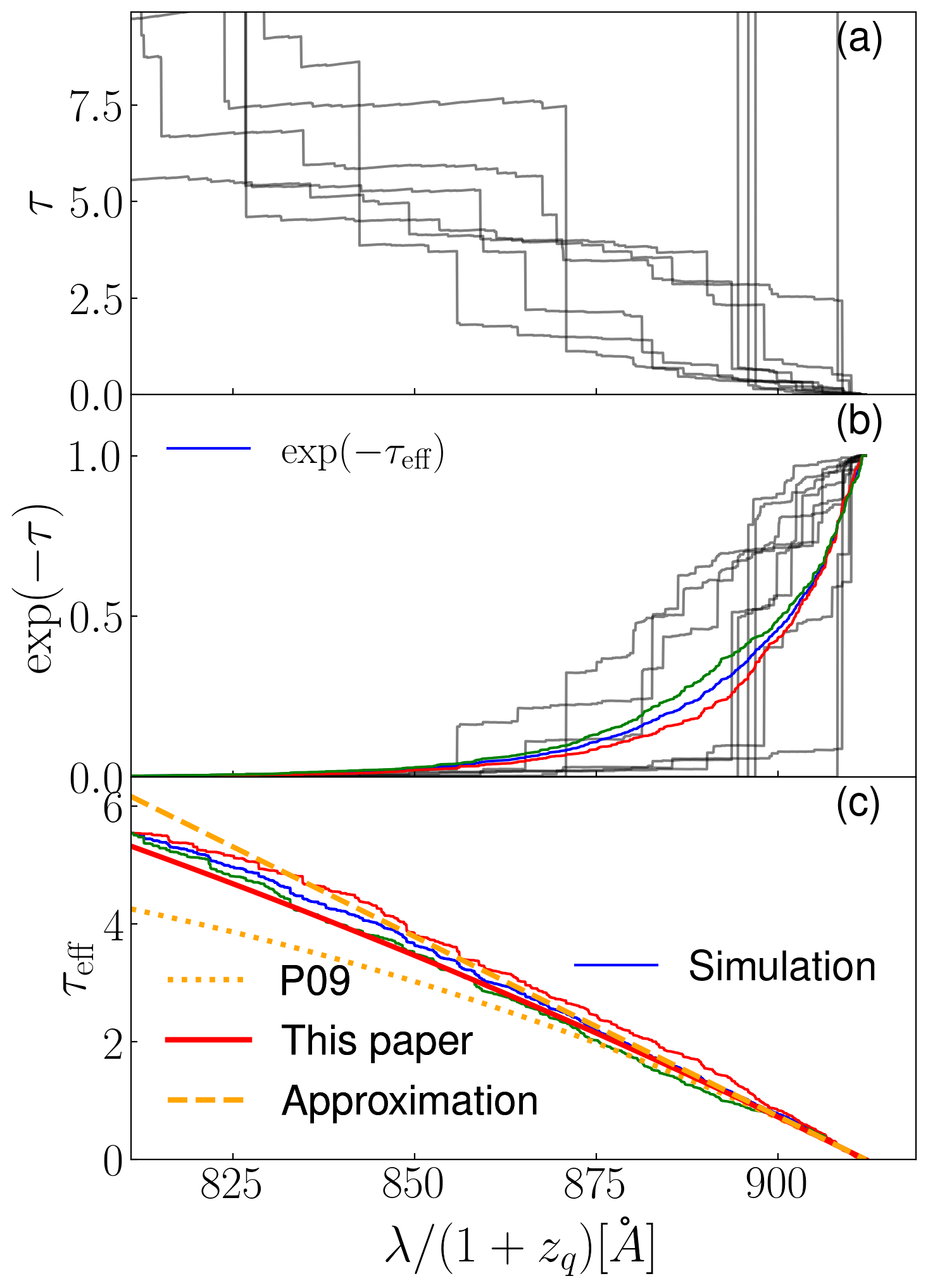}
\caption{Mock spectra and effective optical depths as a function of rest-wavelength, $\lambda_{\rm rest}\equiv \lambda/(1+z_q)$.  {\em Top panel:} Lyman-limit optical depth for 10 mock spectra generated using Poisson-distributed absorption lines. {\em Central panel:} corresponding transmission $\exp(-\tau)$ for these spectra ({\em black thin lines}) and mean transmission for 200 realisations ({\em blue solid line}). The wavy nature of $\exp(-\tau_{\rm eff})$ reflects the relatively large spectrum-to-spectrum variations, a consequence of the relatively low number density of strong absorbers that dominate the optical depth. {\em Lower panel:} effective optical depth for the simulated spectra ({\em blue thin solid line}), the approximation from \protect\cite{Proc09opacity} ({{\em yellow dotted line} labelled \lq P09\rq)}, Eq.~(\ref{eq:taueff1}), and the approximation in this paper ({\em red line}), Eq.~(\ref{eq:taueff2}) with $b_S=0$. {\em Red} and {\em green thin solid lines} in panels (a) and (b) show the simulation estimates using 100 (rather than 200) realisations. Numerically, we set $z_q=6$, $\X=0.607$ is kept constant, integrated the \cddf\ from $\log N_{\rm HI}[{\rm cm}^{-2}]=14\to 22$ in steps of 0.025~dex, and used integration steps of $0.05$~\AA\ in $\lambda_{\rm rest}$.
}
\label{fig:MFP2}
\end{figure}

An intervening absorber with column $N_{\rm HI}\geq 10^{17.2}{\rm cm}^{-2}$ imprints an absorption edge in the spectrum of a quasar at wavelengths $\lambda\leq \lambda_{\rm th}\approx 912.1$~\AA\ in the rest frame of the absorber. Because the photo-ionization cross-section falls $\propto \lambda^{3}$, the optical depth due to such an absorber decreases at lower $\lambda$. However, a second intervening absorber at lower redshift may introduce another absorption edge, which will increase the optical depth again. The total optical depth\footnote{Absorption may also be due to other lines of hydrogen or indeed lines from other elements. We will ignore these in this section.} below $\lambda_{\rm th}$ in the rest-frame of the quasar is therefore a balance between the fall in $\tau$ of any individual \lls\ and the increase in $\tau$ due to the increase in the number of intervening \lls. 

In this section, we use our expression for the evolution of the \cddf\ to compute $\tau_{\rm eff}(\lambda_{\rm rest}, z_q)$ - the effective optical depth as a function of rest wavelength, $\lambda_{\rm rest}$, for quasars with redshift $z_q$. The shape of this curve depends on \L, and \cite{Proc09opacity} stacked \qso\ spectra in bins of $z_q$ to measure \L$(z)$. They argued that this method has the advantage that it determines \L\ without the need to measure the \cddf\ in the regime of \lls's where it is especially hard to determine the column density of these saturated lines. Here we will show that the actual shape of $\tau_{\rm eff}$ also depends on the \cddf, so inferring \L\ still requires making assumptions on the shape of the \cddf\ in the regime of \lls's.

\cite{Proc09opacity} model $\tau_{\rm eff}(\lambda_{\rm rest}, z_q)$ as\footnote{This is Eq.~(6) of \cite{Proc09opacity}, setting their redshift-dependent opacity $\tilde\kappa_{912}(z')\to
\tilde\kappa_{912}(z_q)=\L^{-1}(z_q)$, the proper attenuation length at redshift $z_q$, and then converting $\L(z_q)\to\X$.}
\begin{align}
\tau_{\rm eff}(\lambda_{\rm rest}, z_q) &=  \frac{2}{9}\,\frac{(1+z_q)^{3/2}}{\X(z_q)\,\Omega_m^{1/2}}\,
\,\left(\frac{\lambda_{\rm rest}}{\lambda_{\rm th}}\right)^{-3/2}
\,\left[1-\left(\frac{\lambda_{\rm rest}}{\lambda_{\rm th}}\right)^{9/2}\right]\,.
\label{eq:X09}
\end{align}
Here, $\lambda_{\rm rest}$ is the wavelength {\em in the rest-frame of the quasar}, {\em i.e.} the observed wavelength is $\lambda_{\rm rest}(1+z_q)$. 
To derive this expression, \cite{Proc09opacity} assume that the effective opacity is of the form
\begin{align}
\kappa_{912}(z', z_q, \lambda_{\rm rest}) \equiv \frac{d\tau_{\rm eff}(z', \lambda_{\rm rest})}{dl}\approx \tilde\kappa_{912}(z_q)\,\left[\frac{\lambda_{\rm rest}(1+z')}{\lambda_{\rm th}(1+z_q)}\right]^3\,,
\end{align}
where they argue that the wavelength dependence is approximate, and further assume that $\tilde\kappa_{912}(z')$ is approximately constant over the small wavelength range studied, so that it can be evaluated at $z'\to z_q$. The attenuation length at the mean redshift of the sample of \qso's is determined by fitting the data to this model. For wavelengths close to $\lambda_{\rm th}$ we find
\begin{eqnarray}
    \tau_{\rm eff}(\lambda_{\rm rest}\lesssim \lambda_{\rm th}, z_q)&\approx& \frac{(1+z_q)^{3/2}}{\X(z_q)\,\Omega_m^{1/2}}\,
    \,\left(1-\frac{\lambda_{\rm rest}}{\lambda_{\rm th}}\right)\,.
    \label{eq:taueff1_approx}
\end{eqnarray}

Our own, slightly different, derivation goes as follows. The effective optical depth measured by an observer at redshift $z_o$ at wavelength $\lambda_o$ in a stack of \qso\ spectra with emission redshift $z_q$, is
\begin{align}
    \tau_{\rm eff}(\lambda_o, z_o, z_q)&=\int_{z_{\rm S}}^{z_q}\,dz'\,[1+b_S\,b_{\rm LLS}\xi(z', z_q)]\nonumber\\
    &\times \int_0^\infty\,dN_{\rm HI}\,\frac{dX}{dz'}\,f(N_{\rm HI}, z')\,\left[1-\exp(-\tau)\right]\,,\nonumber\\
    \label{eq:taueff1}
\end{align}
for $\lambda_o\le \lambda_{\rm th}(1+z_q)/(1+z_o)$ and zero otherwise. To see why, notice that the inner integral sums the contribution to $\tau_{\rm eff}$ over column density whereas the outer integral sums over all intervening absorbers that cause bound-free absorption at wavelength $\lambda_o$. 
For wavelengths close to $\lambda_{\rm th}$ in the rest frame of the quasar, only absorbers with redshift close to $z_q$ contribute to the integral over $z$, because the photon's wavelength will be redshifted below the Lyman limit when $z$ is too low. For wavelengths shorter than $\lambda_{\rm th}$ in the rest frame of the observer, all absorbers with $z_o\lesssim z\lesssim z_q$ contribute to the absorption.
The lower limit to the integral over $z'$ is therefore
\begin{align}
    z_{\rm S} = {\rm max}\left[z_o, \frac{\lambda_o}{\lambda_{\rm th}}(1+z_o)-1\right]\,.
    \label{eq:zs}
\end{align}

The quantity $\tau$ in Eq.~(\ref{eq:taueff1}) is the optical depth (and not the effective optical depth) measured by the observer (at redshift $z_o$) at wavelength $\lambda_o$ due to an absorber with column density $N_{\rm HI}$ at redshift $z'$ (with $z_o\le z'\le z_q$), $\tau=\sigma\times N_{\rm HI}$. The photo-ionization cross section, $\sigma$, is a function of the ratio of the Lyman-limit wavelength $\lambda_{\rm th}\approx 912.1$~\AA, over the wavelength of the photon in the rest frame of the absorber. The latter wavelength is $\lambda_o\times (1+z_o)/(1+z')$. We will write the wavelength  dependence of $\sigma$ as \citep[e.g.][]{Verner96}
\begin{align}
    \sigma(\lambda) &
    =\sigma_{\rm th}\times
    \left(\frac{\lambda_{\rm th}}{\lambda}\right)^{-3}
    \equiv \sigma_{\rm th}\times 
    s(\frac{\lambda_{\rm th}}{\lambda})\,,
    \label{eq:sigma_nu}
\end{align}
with the function $s$ encoding the wavelength dependence. Substituting this in the expression for $\tau$ then yields
\begin{align}
\tau(\lambda_o, z_o, N_{\rm HI}, z') = \sigma_{\rm th}\times N_{\rm HI}\,\times 
s\left(\frac{\lambda_{\rm th}(1+z')}{\lambda_o(1+z_o)}\right)\,.
\label{eq:tau_zp}
\end{align}

We now change the integration variable in the inner integral of Eq.~(\ref{eq:taueff1}) from $N_{\rm HI}\to\tau$, using Eq.~(\ref{eq:tau_zp}). This allows us to write Eq.~(\ref{eq:taueff1}) in terms of \X\ evaluated at $\lambda_{\rm th}$ and redshift $z'$ as
\begin{align}
\tau_{\rm eff}(\lambda_o, z_o, z_q)&=\int_{z_{\rm S}}^{z_q}\,dz'\,
\frac{dX/dz'}{\X(z')}\,\left[1+b_S\,b_{\rm LLS}\xi(z', z_q)\right]\nonumber\\
&\times \left[s\left(\frac{\lambda_{\rm th}\,(1+z')}{\lambda_o\,(1+z_o)}\right)\right]^{2/3}\,.
\label{eq:taueff2}
\end{align}

We can compare this (more general) expression to the special case considered by \cite{Proc09opacity} by setting $z_o=0$ and $\lambda=\lambda_o$ (since we are the observer), and making the same four approximations that resulted in Eq.~(\ref{eq:taueff1_approx}): ({\em i}) replace $\X(z')\to\X(z_q)$ ({\em i.e.} assume that the absorption distance does not change appreciably over the small redshift interval), ({\em ii}) take the cross-section $s(x)=x^{3}$, ({\em iii}) neglect clustering of absorbers ({\em i.e.} take $b_S\times b_{\rm LLS}=0$), and finally ({\em iv}) obtain an expression for sufficiently long wavelengths so that we can take $z_{\rm S}=z_o$. This yields
\begin{align}
   \tau_{\rm eff}(\lambda=\lambda_o, z_o=0, z_q) &\approx  \frac{2}{\X(z_q)\Omega_m^{1/2}}\,\left(\frac{\lambda}{\lambda_{\rm th}}\right)^{3/2}\nonumber\\
   &\times \left\{1-\left[\frac{\lambda}{\lambda_{\rm th}\,(1+z_q)}\right]^{1/2}\right\}\,,
   \label{eq:taueff3}
\end{align}
The limit of this expression for $\lambda\to \lambda_{\rm th}(1+z_q)$ is identical to Eq.~(\ref{eq:taueff1_approx}), that is, our alternative expression Eq.~(\ref{eq:taueff2}) is identical to that of \cite{Proc09opacity} close to the quasar (when neglecting bias). However, they differ further away from the \qso. The reason for the difference becomes clear when looking at Eq.~(3) of \cite{Proc09opacity}, where it is assumed that the \lq opacity\rq\
$\kappa\propto f(N_{\rm HI})\,\exp(-\tau)\propto \lambda^{3}$, whereas in our case the scaling is $\propto s^{2/3}\propto\lambda^2$ in the case of $s(x)\propto x^{3}$. We note that ({\em i}) the dependence on wavelength depends on the {\em slope} of the assumed \cddf\ (which is $5/3$ in our model), and ({\em ii}) the scaling $s(x)\propto x^{3}$ is only approximately valid, and it would be better to use a more accurate expression for the photo-ionization cross section \citep[e.g.][]{Verner96}. A final difference in our derivation compared to that of \cite{Proc09opacity} is that we assume that \X\ is approximately constant, which is not the same as assuming that $\tilde\kappa_{912}(z')\propto (1+z')^3/X_{\rm eff}(z')$ is constant over the relevant redshift interval ({\em i.e.} opacity is not a co-moving quantity).

To test our expression, we generate mock absorption spectra as follows.
Choosing a value for $z_q$ and assuming that the \cddf\ is of the form $f(N_{\rm HI})=f_0\times (10^{17.2}{\rm cm}^{-2}/N_{\rm HI})^{5/3}$ for some amplitude $f_0$, we generate the optical depth as a function of wavelength of the form
\begin{equation}
    \tau(\lambda, z_q) = \sum_{z=z_{912}}^{z_q}\,\sum_{N_{\rm HI}=0}^\infty
    N(N_{\rm HI}, z)\,\tau(\lambda, N_{\rm HI}, z)\,,
    \label{eq:tau_sim}
\end{equation}
where $\tau$ is the optical depth at wavelength $\lambda$ due to an absorber with column density $N_{\rm HI}$ at redshift $z$, taken from Eq.~(\ref{eq:tau_zp}), and $N(N_{\rm HI}, z)$ is the Poisson distributed number of lines with \cddf\ $f(N_{\rm HI})$. The mean of this Poisson distribution is $({dX/dz})\,f(N_{\rm HI})\,\Delta N_{\rm HI}\,\Delta z$, where $\Delta N_{\rm HI}$ and $\Delta z$ are the steps in the sums over column density and redshift in Eq.~(\ref{eq:tau_sim}). For a given realisation of $\tau(\lambda, z)$, we can compute the transmission, $\exp(-\tau)$, and averaging over many realisation the effective optical depth, $\langle\exp(-\tau)\rangle\equiv \exp(-\tau_{\rm eff})$. The results of this exercise are summarised in Fig.~\ref{fig:MFP2}, which shows that within our assumed approximations, Eq.~(\ref{eq:taueff3}) (solid red line) reproduces $\tau_{\rm eff}(\lambda_{\rm rest})$ from the simulation very well, with Eq.~(\ref{eq:taueff3}) (dashed yellow line) capturing correctly the gradient of this curve close to $\lambda_{\rm th}$. The original expression 
Eq.~(\ref{eq:taueff1_approx}) from \cite{Proc09opacity} ({yellow dotted line}) falls a little  below the simulated results (blue line).

Finally, we note that biasing and the \qso's proximity effect ({\em i.e.} the fact that the \qso\ itself emits ionising radiation) will likely play an increasingly important role at higher $z$. We could account for the proximity effect by replacing $\Gamma\to \Gamma_0+\Gamma_{\rm q}(z)$ in the expression for \X\ of Eq.~(\ref{eq:X}), where $\Gamma_{\rm q}(z)$ is the photo-ionization rate at redshift $z$ due to the \qso\ itself.

\subsection{{The transition to a transparent Universe}}
\begin{figure}
\includegraphics[width={0.95\columnwidth}]{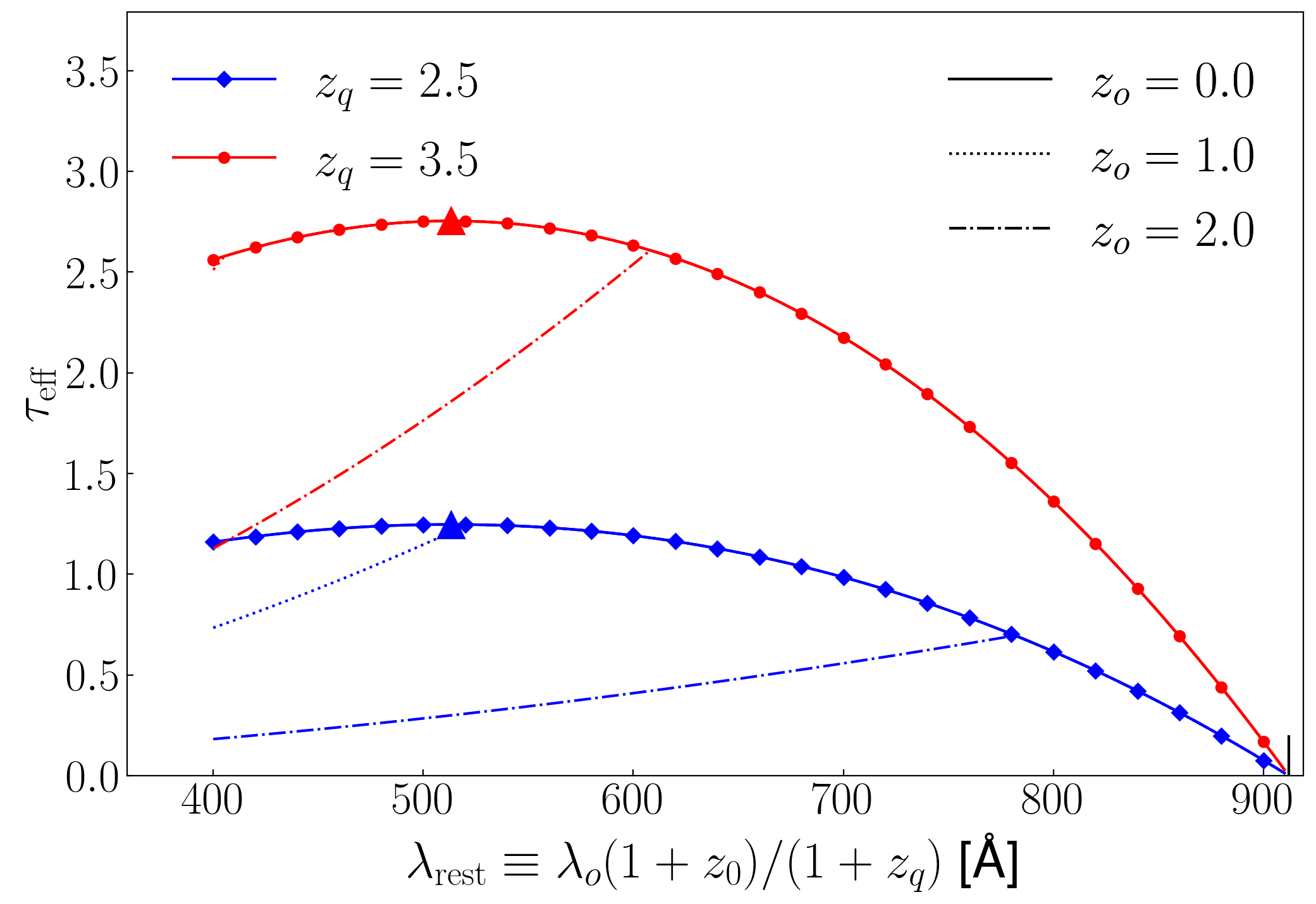}
\caption{Effective optical depth, $\tau_{\rm eff}$, as a function of wavelength, $\lambda_{\rm rest}$, measured in the rest-frame of the \qso, as given by Eqs.~(\ref{eq:bo1}-\ref{eq:bo2}). The {\em red line connecting filled circles} and the {\it blue line connecting filled diamonds} correspond to \qso\ redshifts of $z_q=2.5$ and $3.5$, the observer's redshifts are 0, 1 and 2 for {\it solid, dotted and dot-dashed lines}. The location and value where $\tau_{\rm eff}$ reaches a maximum, given by Eq.(\ref{eq:tmax}) are indicated by a large filled triangle. The Lyman limit rest wavelength is indicated by a small vertical line.}
\label{fig:Break1}
\end{figure}
\begin{figure}
\includegraphics[width={0.95\columnwidth}]{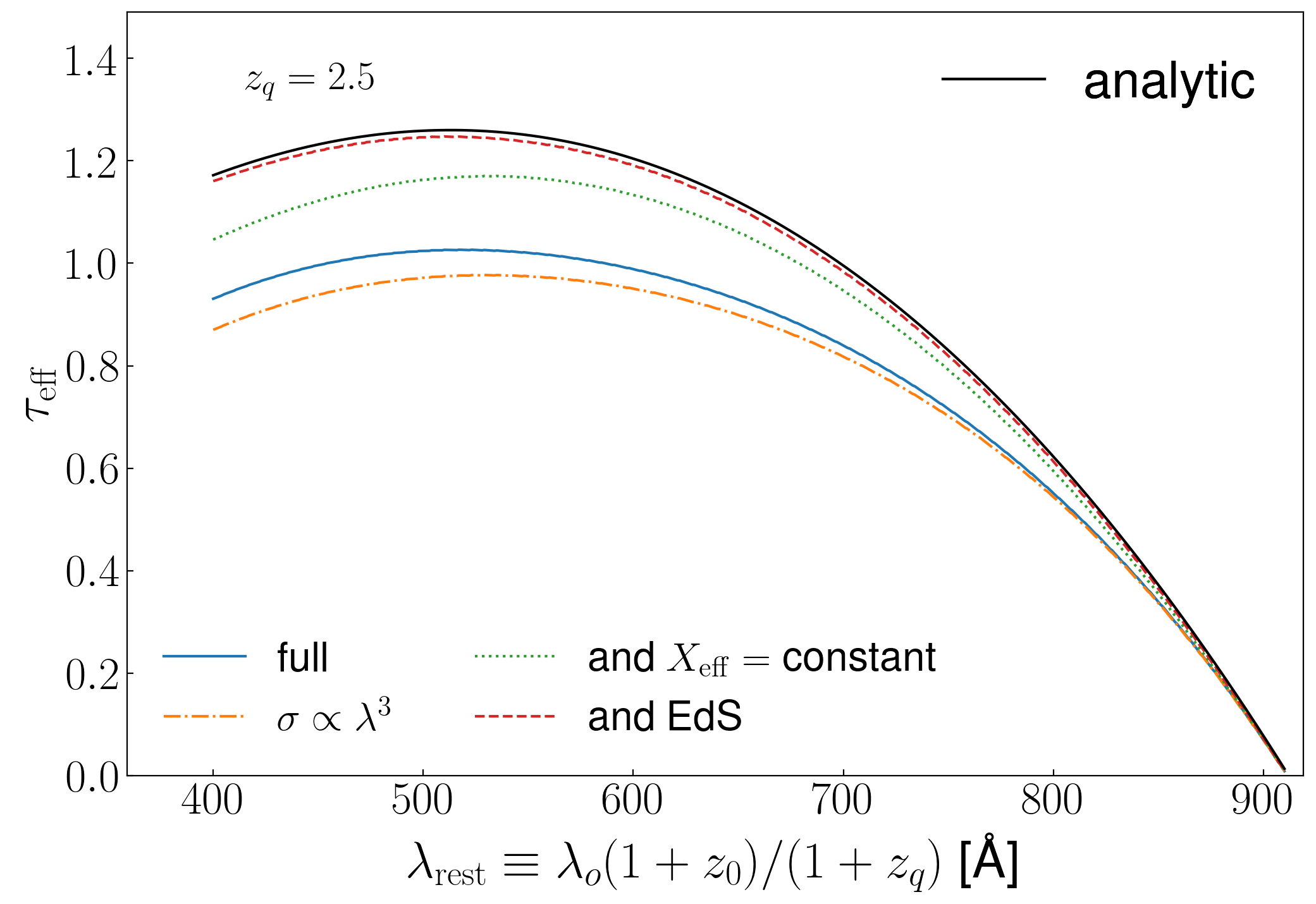}
\caption{Same as Fig.~\ref{fig:Break1} but for a \qso\ at $z_q=2.5$ and an observer at $z_o=0$. The {\em solid blue line} is obtained by numerically integrating Eq.~(\ref{eq:taueff2}). The other line styles are successive approximations: the {\em dot-dashed orange line} takes the photo-ionization cross section to be $\propto \lambda^{-3}$ rather than the more accurate expression from \protect\cite{Verner96}, 
the {\em dotted green line} in addition takes $X_{\rm eff}(z)$ constant at its value at $z=z_q$, and the {\em dashed red line} in addition assumes an EdS universe, $H(z)=H_0\,\Omega_m^{1/2}(1+z)^{3/2}$. The {\em solid black line}
is the analytical expression from Eqs.~(\ref{eq:bo1}-\ref{eq:bo2}) which makes the same approximations as the {\em dashed red line}; it has been off-set vertically by a factor 1.01 to avoid complete overlap with that line.
}
\label{fig:Break2}
\end{figure}

To interpret the general expression for $\tau_{\rm eff}$ of Eq.(\ref{eq:taueff2}) qualitatively, it is useful to make the following approximations,
({\em i}) use the high-$z$ expression for the Hubble constant, ({\em ii}) assume that the hydrogen photo-ionization cross section has 
wavelength dependence $\sigma\propto \lambda^{3}$, ({\em iii}) take the absorption distance $X_{\rm eff}(z')$ in the expression to be constant at its value for $z=z_q$, and ({\em iv}) neglect clustering ($b_S\times b_{\rm LLS}\to 0$). This yields the following analytical expressions,

\begin{align}
   \tau_{\rm eff}(\lambda_{\rm rest}, z_o, z_q) &\approx  \frac{2}{\X(z_q)\Omega_m^{1/2}}\,\left[\frac{\lambda_{\rm rest}(1+z_q)}{\lambda_{\rm th}}\right]^{3/2}\nonumber\\
   &\times \,\left[1-\left(\frac{\lambda_{\rm rest}}{\lambda_{\rm th}}\right)^{1/2}\right]\,,
   \label{eq:bo1}
\end{align}
for $\lambda_{\rm th}(1+z_o)/(1+z_q)\le \lambda_{\rm rest}\le \lambda_{\rm th}$, and
\begin{align}
   \tau_{\rm eff}(\lambda_{\rm rest}, z_o, z_q) &\approx  \frac{2}{\X(z_q)\Omega_m^{1/2}}\,\left[\frac{\lambda_{\rm rest}(1+z_q)}{\lambda_{\rm th}}\right]^{2}\nonumber\\
   &\times \frac{1}{(1+z_o)^{1/2}}\left[1-\left(\frac{1+z_o}{1+z_q}\right)^{1/2}\right]\,,
   \label{eq:bo2}
\end{align}
for $\lambda_{\rm rest}\le \lambda_{\rm th}(1+z_o)/(1+z_q)$, where as before,
$\lambda_{\rm rest}$ is the photon's wavelength in the rest frame of the \qso.
The first expression has a maximum optical depth, $\tau_{\rm eff, max}$ which occurs at a rest-wavelength $\lambda_{\rm max}$, given by
\begin{align}
\lambda_{\rm rest, max} &=\left(\frac{3}{4}\right)^2\,\lambda_{\rm th}\nonumber\\
\tau_{\rm eff, max} &\approx 0.105\times \frac{2}{\X(z_q)\Omega_m^{1/2}}\,(1+z_q)^{3/2}\,.
\label{eq:tmax}
\end{align}
The motivation for computing these expressions for observers at different redshifts - and not just for $z_o=0$ - is that $\tau_{\rm eff}(\lambda_{\rm rest}, z_o, z_q)$ can be used to compute the photo-ionization rate at redshift $z_o$ due to a \qso\ at higher $z$.

The resulting run of optical depth with wavelength is plotted in Fig.~\ref{fig:Break1} for two \qso\ redshifts ($z_q=2.5$ and 3.5) and three observer redshifts ($z_o=0$, 1 and 2).  The shape of these curves can be understood as follows. Photons with rest wavelength close to $\lambda_{\rm th}$ can only be absorbed by absorbers close to the \qso\ before they redshift below the Lyman limit of intervening neutral gas. Therefore the redshift path where an absorber affects the photon lengthens with decreasing $\lambda_{\rm rest}$: this is why $\tau_{\rm eff}$ initially increases with decreasing wavelength. There are two reasons why $\tau_{\rm eff}$ eventually starts to decrease again with decreasing  $\lambda_{\rm rest}$. Firstly, once a photon's wavelength becomes smaller than $\lambda_{\rm th}$ in the rest-frame of the observer, $z_S\to z_o$, the redshift range that contains absorbers, ceases to lengthen. The optical depth then drops because the photo-ionization cross section drops, and $\tau_{\rm eff}$ is given by Eq.~(\ref{eq:bo2}) rather than Eq.~(\ref{eq:bo1}).
This sudden change is illustrated by the dotted and dot-dashed lines that branch away from the solid line in the Figure. Secondly, $\tau_{\rm eff}$ starts to decrease once $\lambda_{\rm rest}\le (3/4)^2\lambda_{\rm th}$, even when $\tau_{\rm eff}$ is described by Eq.~(\ref{eq:bo1}). This occurs because
the tension between $\tau_{\rm eff}$ increasing due to the increasing redshift path (due to the factor $1-(\lambda_{\rm rest}/\lambda_{\rm th})^{1/2}$) and $\tau_{\rm eff}$ decreasing due to the decreasing photo-ionization cross-section
(the factor $(\lambda_{\rm rest}/\lambda_{\rm th})^{3/2}$) is eventually decided in favour of the latter process. We note that the decrease in $\tau_{\rm eff}$, in this case, is {\em not} due to the decrease in the co-moving number of absorbers at lower redshifts, since in the approximation that leads to these equations we have kept $X_{\rm eff}$ {\em constant}.

Although Eqs.~(\ref{eq:bo1}-\ref{eq:bo2}) are useful for describing the qualitative behaviour of $\tau_{\rm eff}$, they are not particularly accurate because the approximations made in deriving them from  Eq.~(\ref{eq:taueff2}) are not very accurate, as we illustrate in Fig.~\ref{fig:Break2}. Of the various approximations made, we see that accounting for the evolution of $X_{\rm eff}$ has the largest impact. Indeed, if we allow $X_{\rm eff}$ to increase with decreasing $z$ using Eq.~(\ref{eq:xeff}), the value of $\tau_{\rm max}$ is reduced by about 20 per cent for this particular choice of $z_q$. 

Finally, we note that $\tau_{\rm eff}$ reaches a maximum value of $\sim 1$ for a \qso\ at redshift $z_q\sim 2.5$. This means that below a redshift of 2.5, most \qso's\ contribute to ionizing neutral hydrogen atoms at all lower $z$, {{\em i.e.} the Universe becomes \lq transparent\rq\ to ionizing radiation}. Indeed, unless the atoms are in a self-shielded region, intervening absorbers typically decrease the ionizing flux by less than a factor of $1/e$. \cite{Madau99} referred to this epoch as \lq breakthrough\rq. Their value of the breakthrough redshift of $\sim 1.6$ is lower than our value of $2.5$. The reason is that they assumed that $\X\propto (1+z)^{-3/2}$ with a normalization set by the \cddf\ at $z=3$; they also use a different slope for the \cddf.

\section{Summary and conclusions}
We presented an analytical model for the column-density distribution function (\cddf) of hydrogen absorption lines along a sight line piercing the intergalactic medium (\igm; Eq.~\ref{eq:f17}). The model assumes that cosmic gas in dark matter halos follows a power law distribution in density, $\rho(R)\propto R^{-2}$, and is photo-ionized by an evolving radiation background with amplitude $\Gamma_0(z)\equiv \Gamma_{-12}\times 10^{-12}{\rm s}^{-1}$, as computed by \cite{Haar12UVbackground}. The resulting \cddf\ reproduces well the observed \cddf\ at redshift $z\sim 3$ for hydrogen column densities $\log N_{\rm HI}[{\rm cm}^{-2}]$ in the range [$14\to 16$], and [$20\to 22$], where the \cddf\ is well-measured (Fig.\ref{fig:MF1}). The analytical expression for the \cddf\ contains one free parameter, $f_{\rm gas}$, which is of order unity, and some extra parameters such as the temperature of the gas for which we use observed values. The evolution of the model's \cddf\ is due to ({\em i}) the evolution of $\Gamma_{-12}(z)$, ({\em ii}) the evolution of $M_{\rm crit}$, which is the halo mass below which halos lose their gas due to photo-heating by the radiation background, ({\em iii}) the dependence of the virial temperature of a halo of given mass on $z$, and finally, and to a lesser extent, ({\em iv}) the evolution of the halo mass function. Our model builds on that of \cite{Theuns21}, as well as earlier models by \cite{Mira00reion} and 
\cite{Muno16flatgamma}.

We then use the model to compute the evolution of the attenuation length of ionizing photons, \L\ (see Eq.~\ref{eq:lambda}). The evolution of 
\L\ is dominated by cosmological expansion, while the co-moving evolution is due to the evolution of the \cddf. We find that the model's evolution of \L\ agrees very well with the observed evolution in the redshift range $z=2\to 5$, but not for $z>5$  where the data evolve much faster than the model (Fig.\ref{fig:MF1}). Even though the model reproduces the value of \L\ at $z\sim 3$ very well, it underestimates the number of absorption lines with $\log N_{\rm HI}({\rm cm}^{-2})>17.5$ by about a factor of two (see Fig.~\ref{fig:lX}).

Since absorption lines occur when a sight line 
intersects a halo in our model, we can relate the clustering of halos to that of the corresponding absorbers. The bias of Lyman-limit systems (\lls's) is $\sim 1.5$ at $z=2$, increasing to $b\sim 2.6$ at $z=6$ (Fig.~\ref{fig:Bias}). At first surprising, we find that the bias of damped Lyman-$\alpha$ systems (\dla's) with $\log N_{\rm HI}({\rm cm}^{-2})=20.3$ is {\em lower} than that of \lls . The reason is that self-shielding - which causes the transition from highly ionized \lls's to mostly neutral \dla's - sets in at lower column density in lower mass halos - and such halos are less biased. At even higher columns, the bias of \dla's increases rapidly with increasing $N_{\rm HI}$.

We account for clustering between absorbers and quasars, assuming that quasars inhabit dark matter halos with masses $M_h\approx 10^{12-13}{\rm M}_\odot$ (Fig.~\ref{fig:MFP-bias}), and reach the following conclusions. Bias has little effect on the value of \L\ inferred from quasar spectra below $z\sim 4$. However, the rapid increase in quasar host bias above this redshift leads to a corresponding rapid decrease in the value of \L\ inferred from analysing quasar spectra, and this brings the model's evolution of \L\ into line with the observations, also at $z\sim 6$. It is important to realise that this finding has potential implications when studying the tail-end of reionization at $z\sim 6$: the value of \L\ {\em measured in quasar spectra} is generally less (by almost an order of magnitude) than the value of \L\ in the \igm. This makes it harder for quasars to ionize the \igm, since they are surrounded by many more absorbers than galaxies: $\Gamma\propto \L$ so that galaxies contribute more to the ionizing background than quasars, {\em even in the case that both population had the same emissivity}\footnote{Note that the bias we compute is the two-halo term: the absorbers we consider inhabit a different halo from the source. There may be an additional effect from associated absorbers. We also note that we have not accounted for other proximity effects.}. We use our model to calculate the statistics of the attenuation length for rays of a given length in \S \ref{sect:biaslambda} (see Fig.~\ref{fig:MFP}). We find that the distribution of mean transmissions for rays with a given length of the order of \L\ has a long tail to very large values of $\tau_{\rm eff}$, a consequence of the fact that the absorption is dominated by relatively strong absorbers which are rare.

In the final section \S \ref{sect:measuring}, we use the model to compute the mean transmission, $T(\lambda_{\rm rest}, z_q)$,
due to Lyman-limit absorption (where $\lambda_{\rm rest}$ is wavelength in the \qso's rest frame and $z_q$ the redshift of the quasar in which $T$ is measured). We relate $T$ to the amplitude and slope of the \cddf\ around column densities $\sim 10^{17.2}{\rm cm}^{-2}$, and examine how it is affected by various commonly made simplifications. Our general expression, Eq.~(\ref{eq:taueff2}), reduces to that derived by \cite{Proc09opacity} for wavelengths close to 912\AA\ in the rest-frame of the quasar, but differs at shorter wavelengths. We find that the minimum transmission $T$ stays above $e^{-1}$ (i.e. the corresponding effective optical depth remains below 1) on average when $z_q\lesssim 2.5$, which is, therefore, the earliest redshift below which the Universe becomes transparent to ionizing photons.

This paper shows that a simple model for gas in halos accurately predicts the evolution of the \cddf\ and that of the associated attenuation length. The model also allows us to account for bias and clustering. Of course, our analytical model is not as accurate nor as realistic as numerical simulations, but it illustrates well the dominant properties of halos and the \igm\ that give rise to the observables. Several aspects of the model could be further improved. These include accounting for scatter in the gas properties for halos of a given mass and deviations from spherical symmetry, and a more accurate treatment of the temperature of the absorbing gas. The model assumes that the density profile of the gas\footnote{It would be easy to redo the calculations for another assumed power-law.} is $\rho(R)\propto R^{-2}$, and it would be worthwhile examining why this assumption works so well. In the model, the majority of strong absorbers occur in the outskirts of dark matter halos, with some smaller fractions occurring outside the virial radius of the halo. This is consistent with the observation that such absorbers also correlate strongly with the presence of nearby galaxies \citep{Lofthouse23}. This also implies that the sources of the ionizing photons inhabit the same dark matter halos as the sinks. It would be worth exploring whether this correlation can be accounted for \citep[see e.g.][]{Muno16flatgamma}, rather than combining a model for the absorbers with the \cite{Haar12UVbackground} model for the ionizing background as we did here.

\section*{ACKNOWLEDGEMENTS}
We thank the referee for their comments and suggestions, which improved the paper. {\sc TKC} is supported by the E. Margaret Burbidge Prize Postdoctoral Fellowship from the Brinson Foundation at the Departments of Astronomy and Astrophysics at the University of Chicago. We thank S.~Morris and M.~Fumagalli for comments on an earlier draft. This work was supported by the Science and Technology Facilities Council (STFC) astronomy consolidated grants ST/P000541/1
and ST/T000244/1. This work used the DiRAC@Durham facility managed by the Institute for Computational Cosmology on behalf of the \href{www.dirac.ac.uk}{STFC DiRAC HPC Facility}. The equipment was funded by BEIS capital funding via STFC capital grants ST/K00042X/1, ST/P002293/1, ST/R002371/1 and ST/S002502/1, Durham University and STFC operations grant ST/R000832/1. DiRAC is part of the UK's National e-Infrastructure. We used the {\sc matplotlib} \citep{Hunt07matplotlib}, {\sc numpy} \citep{vand11numpy}, {\sc scipy} \citep{Jone01scipy} and {\sc colossus} \citep{Diemer18COLOSSUS} python libraries, and the \href{https://ui.adsabs.harvard.edu/}{NASA’s Astrophysics Data System} digital library portal and \href{https://arxiv.org/}{Xarchiv} open-access repository of electronic e-prints. For the purpose of open access, the author has applied a Creative Commons Attribution (CC BY) licence to any Author Accepted Manuscript version arising.

\section*{DATA AVAILABILITY}
This paper does not contain any new data.

\bibliographystyle{mn2e}
\bibliography{mn-jour,mybib}
\appendix
\section{Attenuation length versus mean free path}
\label{sect:appendixB}
We relate the mean free path to the attenuation length due to a distribution of absorbers as follows.
Consider a Poisson distribution of absorbers with mean number density per unit distance $\mu$, all of which have the same optical depth, $\tau_i$. The probability of having more than $N'$ such absorbers in a distance $L$, is given by
\begin{equation}
    {\cal P}(>N'|L)=1-{\cal P}(N=0, 1, 2, \cdots N'|L)
                  =1-\sum_{N=0}^{N'}\,{\mathbb P}(N' | N \mu L)\,,
\end{equation}
where ${\mathbb P}(x|y)\equiv y^x\,\exp(-y)/x!$ is the Poisson distribution. The probability of reaching $N'$ absorbers after travelling a distance between $L$ and $L+dL$ is the derivative of this cumulative distribution with respect to $L$, which is the Gamma distribution
\begin{equation}
    {\cal P}(L)=\frac{\mu\,\exp(-\mu L)\,(\mu L)^{N'-1}}{(N'-1)!}\,.
    \label{eq:pL}
\end{equation}
This is a well-known result in statistics.

We define the free path of a photon to be the distance it travelled before encountering an optical depth $\tau>1$. In our case, this corresponds to encountering more than $N'=1/\tau_i$ absorbers. The \pdf\ of the free path is therefore given by Eq.~(\ref{eq:pL}), provided we set
$N'=1/\tau_i$. The mean value of the free path - {\em i.e.} the mean free path - is then

\begin{equation}
    \lambda=\langle L\rangle = \int_0^\infty {\cal P}(L)\,dL=\frac{(\tau_i^{-1}+1)\,\Gamma(\tau_i^{-1})}{\mu\,(\tau_i^{-1}-1)!} \approx\frac{1}{\tau_i\mu}\,.
    \label{eq:ALMFP}
\end{equation}

On the other hand, the effective optical depth encountered after travelling a distance $L$, is
\begin{equation}
    \tau_{\rm eff}(L)=\mu\,L\,(1-\exp(-\tau_i))\,.
\end{equation}
The attenuation length, $\lambda_{\rm eff}$ - the distance travelled to reach $\tau_{\rm eff}=1$ - is therefore
\begin{equation}
    \lambda_{\rm eff}=\frac{1}{\mu\,\left(1-\exp(-\tau_i)\right)}\,.
    \label{eq:Aatt}
\end{equation}
Comparing Eq.~(\ref{eq:ALMFP}) to Eq.~(\ref{eq:Aatt}) shows that the attenuation length equals the mean free path in the limit of $\tau_i\ll 1$, but for $\tau_i=0.5$, for example, $\lambda=2/\mu$ but $\lambda_{\rm eff}=2.5/\mu$.

Consider now the the case of $\tau_i\to\infty$. In that limit, the free path is the distance travelled up to the first absorber, therefore the \pdf\ of $L$ becomes
\begin{equation}
    {\cal P}(L)=\frac{1}{\mu}{\mathbb P}(0| \mu L)=\frac{1}{\mu}\exp(-\mu\,L)\,,
\end{equation}
so that the mean free path is $\lambda=\langle L\rangle=\mu^{-1}$. The mean transmission 
after a distance $L$ is the fraction of paths that did not encounter an absorber, 
$\exp(-\tau_{\rm eff}(L))={\mathbb P}(0|\mu L)$. Therefore, the attenuation length is
$\lambda_{\rm eff}=\mu^{-1}$ - and hence equals the mean free path.

In conclusion: when absorption is dominated by very strong absorbers (the case of $\tau_i\to\infty$) or in the case of a uniform \igm\ (the case of $\tau_i\to 0$), mean free path and attenuation length have the same numerical value. However, if a significant fraction of the absorption is due to absorbers with optical depth of order unity, then the attenuation length is larger than the mean free path. The latter case applies to Lyman-limit absorption in the \igm. The attenuation length is often and erroneously referred to as mean free path in the literature - which is unfortunate.

\section{The evolution of the number density of lls}
\label{sect:lls}
\begin{figure}
\includegraphics[width={0.95\columnwidth}]{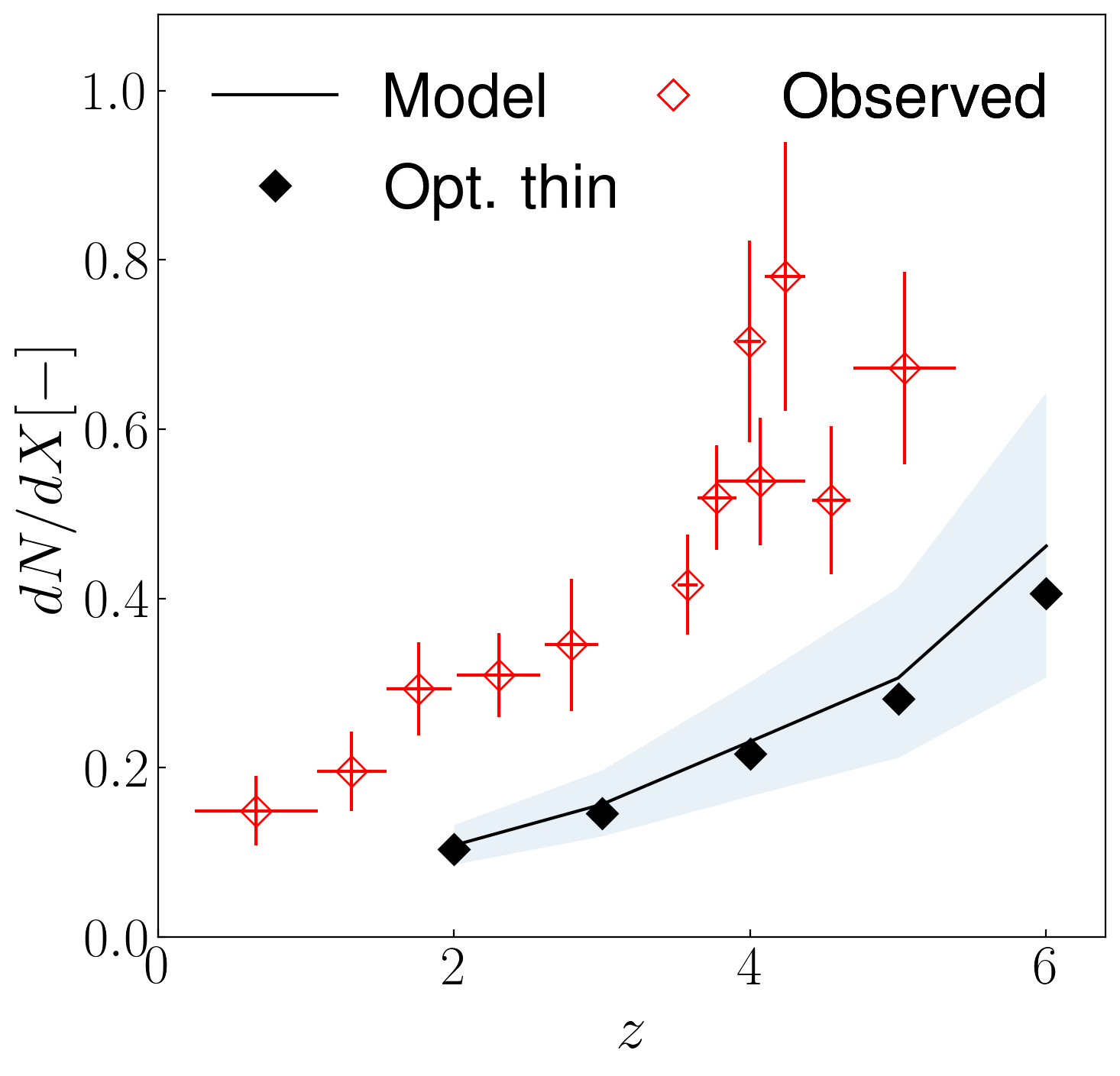}
\caption{Evolution of the mean number density of absorbers with optical depth $\tau>2$. The {\em black solid line} is the evolution predicted by the model of \citetalias{Theuns21}, the {\em grey shading} corresponds to varying the value of the parameter $M_{\rm crit}$ by factors between 1/4 and 4. The {\em black solid diamonds} are the approximate evolution of the model from Eq.~(\ref{eq:dNdX}) using the power-law of the \cddf\ from Eq.~(\ref{eq:fN}). {\em Red symbols with error bars} are the observed data plotted in Fig.~9 of \protect\cite{Crighton19}.  Model and cosmological parameters are as in Fig.~\ref{fig:MF1}. }
\label{fig:lX}
\end{figure}

\cite{Crighton19} review different methods for identifying strong {\sc HI} absorbers 
in {\sc qso} spectra. They then present results of a survey for such absorbers in a homogeneous dataset of 153 {\sc qso} spectra at redshift $z\sim 5$ from the Giant Gemini {\sc gmos} survey \citep{Worseck14}. Combining values from the literature with their own analysis, they present the evolution of the number density of strong absorbers in terms of the co-moving quantity $l(X)$, which is the mean number density of absorbers (with column density larger than some value) per unit co-moving path length, $X$. They count absorbers with $N_{\rm HI}>10^{17.5}{\rm cm}^{-2}$ because these can be identified confidently given the limited signal-to-noise ratio of their data.

Given that $l(X)$ is a number density of absorbers, we prefer to use the notation $dN/dX$, rather than $l(X)$, since $l(X)$ is easily mistaken for a length. Without further ado, we find the following relation between $dN/dX\equiv l(X)$ and the \cddf, where on the second line we substitute the approximate relation of Eq.~(\ref{eq:fN2}) for the \cddf,
\begin{align}
    \frac{dN}{dX}(z) &=\int_{10^{17.5}{\rm cm}^{-2}}^\infty f(N_{\rm HI})\,dN_{\rm HI}\approx 0.2\,
    \frac{f_{17.2}(z)}{\Gamma^{2/3}_{-12}(z)}\,.
    \label{eq:dNdX}
\end{align}
This relation follows from either integrating the \cddf\ of Eq.~(\ref{eq:fN}) from $N_{\rm HI}=10^{17.5}{\rm cm}^{-2}\to\infty$ or directly from Eq.~(\ref{eq:sigma}). The latter route makes it clearer why $dN/dX$ does not depend on the shape of the \cddf\ for $N_{\rm HI}>10^{17.5}{\rm cm}^{-2}$. 

Figure~\ref{fig:lX} compares the evolution predicted by the model to the observations plotted in Fig.~9 of \cite{Crighton19}. The data are compiled from \cite{Ribaudo11,Prochaska10,Omeara13} and \cite{Fumagalli13}, in addition to data from \cite{Crighton19}. We first note that the optically thin approximation (black diamonds) reproduces almost exactly \citetalias{Theuns21}'s model that includes self-shielding (black curve). Both underestimate the observed number density (red diamonds) by a factor $\sim 2$ yet reproduce the observed evolution very well. It is somewhat surprising that the model described so far reproduces \L\ well for $z\leq 5$ as seen in Fig.~\ref{fig:MF1} (right panel) yet it underestimates the number of \lls's with $N_{\rm HI}\ge 10^{17.5}{\rm cm}^{-2}$ by a factor 2. The middle panel of Fig.~\ref{fig:MF1} shows why this is: these higher column density \lls\ actually contribute little to \L.

We venture that scatter in the density distribution around halos may be the main culprit for the underestimate in $dN/dX$ in the model. Indeed, these higher column density systems have by construction an optical depth to ionizing photons of around unity. Consequently, a small increase in total column density may result in an exponential increase in neutral column density due to the onset of self-shielding. The impact of such {\em scatter} on the \cddf\ is substantial: a 0.2~dex Gaussian scatter in $\log N_{\rm HI}$ results in a factor of 2 increase in $dN/dX$ - enough to bring the model in good agreement with the data. Importantly, this exponential dependence on column density mostly affects absorbers around the knee of the \cddf, where the absorbers transition from optically thin to optically thick.
\section{Statistics of $\tau_{\rm eff}$}
\label{sect:AppendixA}
In this Appendix we derive an approximate analytical expression for the {\sc pdf} of $\tau_{\rm eff}$, as discussed in section~\ref{sect:biaslambda}. Our derivation goes as follows.
At sufficiently low $N_{\rm HI}$, the mean number of lines that contribute to $\tau_{\rm eff}$ may be large enough that the central limit theorem is applicable. In that case, the lines are approximately Gaussian distributed (with mean $\langle N\rangle$ and dispersion $\langle N\rangle$). Integrating over $dN_{\rm HI}$, and integrating over $dX$ then corresponds to summing over independently-distributed Gaussian variables. Therefore, the sum is also Gaussian distributed, with mean the sum of the means, and dispersion the sum of the dispersions. We will denote the value of $\tau_{\rm eff}$ due to these low column-density lines by $\tau_{\rm eff, low}$, and its \pdf\ is therefore
\begin{eqnarray}
    {\cal P}_G(\tau_{\rm eff, low}|\mu, \sigma) &=& \frac{1}{(2\pi\sigma^2)^{1/2}}\,\exp(-\frac{(\tau_{\rm eff, low}-\mu)^2}{2\sigma^2})\nonumber\\
    \mu(X) &=& X\,\int_0^{\rm N_{\rm HI, low}}\,f(N_{\rm HI})\,(1-\exp(-\tau))\,dN_{\rm HI}\nonumber\\
    \sigma^2(X) &=& X\,\int_0^{\rm N_{\rm HI, low}}\,f(N_{\rm HI})\,(1-\exp(-\tau))^2\,dN_{\rm HI}\,.\nonumber\\
\end{eqnarray}
We added a subscript \lq $G$\rq\ as a reminder that we assume Gaussian statistics.

We can account for the higher column density absorbers as follows. Let's take $N_{\rm HI, low}=10^{17.2}\,{\rm cm}^{-2}$. In that case, the weighting factor $(1-\exp(-\tau))\approx 1$ for those lines with $N_{\rm HI}\ge N_{\rm HI, low}$. The \pdf\ due to these higher column density lines is now a sum of independently distributed Poisson variables, hence also a Poisson distributed variable\footnote{The reason this works in this approximation is that the weights of each individual Poisson variable are now equal, $1-\exp(-\tau)\to 1$, so now it is a sum rather than a more general linear combination of Poisson variables.}. Denoting the value of $\tau_{\rm eff}$ due to these high column-density lines by $\tau_{\rm eff, high}$, we find that its \pdf\ is given by
\begin{eqnarray}
    {\cal P}_P(\tau_{\rm eff, high}) &=& \mathbb{P}(N|N_P)\nonumber\\
    N_P(X) &=& X\,\int_{N_{\rm N_{\rm HI, low}}}^\infty\,dN_{\rm HI}\,,
\end{eqnarray}
with subscript \lq $P$\rq\ as a reminder that we assume Poisson statistics.

The total effective optical depth is $\tau_{\rm eff, low}+\tau_{\rm eff, high}$, with \pdf\
\begin{equation}
    {\cal P}(\tau_{\rm eff}(X))=\sum_{N=0}^\infty\,
    \mathbb{P}(N|N_P)\,{\cal P}_G(\tau_{\rm eff}-N|\mu, \sigma)\,,
    \label{eq:taueff}
\end{equation}
with $\mu$, $\sigma$ and $N_P$ all proportional to $X$. The mean of this distribution is the sum of the means of 
$\tau_{\rm eff, low}$ and $\tau_{\rm eff, high}$, 
\begin{align}
    \langle\tau_{\rm eff}(X; z)\rangle 
    &=X\,\int_0^1 f(N_{\rm HI})\,\left[1-\exp(-\tau)\right]\,dN_{\rm HI}\nonumber\\
    &+X\,\int_1^\infty f(N_{\rm HI})\,dN_{\rm HI}\nonumber\\
    &\approx  4.2\,X\,\frac{f_{17.2}(z)}{\sigma_{\rm th}}\,,
\end{align}
where we used the power-law approximation to the \cddf\ of Eq.~(\ref{eq:fN2}). The numerical value also shows the limitation of setting $1-\exp(-\tau)\to 1$ for the high column density absorbers. If we had not made that approximation, then $\langle\tau_{\rm eff}(X; z)\rangle =4.02\,X\, {f_{17.2}(z)}/{\sigma_{\rm th}}$.

We can now compute the \pdf\ of the co-moving attenuation length - {\em i.e.} the \pdf\ of $X$ where $\tau_{\rm eff}=1$ - as follows. The fraction of sight lines that reach $\tau_{\rm eff}>1$ for a given value of $X$ is
\begin{align}
    {\cal P}(\tau_{\rm eff}>1|X)&=\int_1^\infty {\cal P}(\tau_{\rm eff}|X)\,d\tau_{\rm eff}\nonumber\\
    &=\frac{1}{2}\sum_{N=0}^\infty \mathbb{P}(N|N_P)\left\{1\pm {\rm Erf}(x)\right\}\nonumber\\
    x&\equiv \frac{1-N-\mu}{(2\sigma^2)^{1/2}}\,.
    \label{eq:taugt1}
\end{align}
Here, ${\rm Erf}$ denotes the error function, and the upper and lower signs applies to the case where $x$ is negative or positive. The fraction of paths that reach $\tau_{\rm eff}$ between $X$ and $X+dX$ follows by taking the derivative with respect to $X$,
\begin{align}
    {\cal P}&(\tau_{\rm eff}=1,X)
    =\frac{d{\cal P}(\tau_{\rm eff}>1|X)}{dX}\nonumber\\
    &=\sum_{N=0}^\infty\left\{ \left(\frac{N}{N_P}-1\right)\,\mathbb{P}(N|N_P)\,\left\{1\pm {\rm Erf}(\mp x)\right\}\,\frac{N_P}{X}\right.\nonumber\\
    &+\left. \mathbb{P}(N|N_P)\,\frac{\exp(-x^2)}{\pi^{1/2}}\left(\frac{x}{2X}+\frac{\mu_G}{(2\sigma^2)^{1/2}X}\right)\right\}\,.
    \label{eq:pdfY}
\end{align}
This is the approximate analytical expression for the probability distribution of the attenuation length that we set out to obtain.
\label{lastpage}
\bsp
\end{document}